\documentclass[11pt]{article}
\pdfoutput=1
\usepackage{amsmath,graphicx,xcolor,amsfonts,float,grffile}
\usepackage[margin=1in]{geometry}
\usepackage{subcaption}
\usepackage{multirow}
\usepackage{url}
\usepackage{pdflscape}
\usepackage{natbib}

\def\bSig\mathbf{\Sigma}

\newcommand\independent{\protect\mathpalette{\protect\independenT}{\perp}}
\def\independenT#1#2{\mathrel{\rlap{$#1#2$}\mkern2mu{#1#2}}}

\title{Estimating Population Average Causal Effects in the Presence of Non-Overlap: The Effect of Natural Gas Compressor Station Exposure on Cancer Mortality}
\author{Rachel C. Nethery$^{1}$, Fabrizia Mealli$^{2}$, and Francesca Dominici$^{1}$ \\
$^{1}$Department of Biostatistics, Harvard T.H. Chan School of Public Health, Boston MA, USA \\
$^{2}$Department of Statistics, Informatics, Applications, University of Florence, Florence, Italy}
\date{}

\begin{document}

\maketitle

\begin{abstract}
Most causal inference studies rely on the assumption of overlap to estimate population or sample average causal effects. When data suffer from non-overlap, estimation of these estimands requires reliance on model specifications, due to poor data support. All existing methods to address non-overlap, such as trimming or down-weighting data in regions of poor data support, change the estimand so that inference cannot be made on the sample or the underlying population. In environmental health research settings, where study results are often intended to influence policy, population-level inference may be critical, and changes in the estimand can diminish the impact of the study results, because estimates may not be representative of effects in the population of interest to policymakers. Researchers may be willing to make additional, minimal modeling assumptions in order to preserve the ability to estimate population average causal effects. We seek to make two contributions on this topic. First, we propose a flexible, data-driven definition of propensity score overlap and non-overlap regions. Second, we develop a novel Bayesian framework to estimate population average causal effects with minor model dependence and appropriately large uncertainties in the presence of non-overlap and causal effect heterogeneity. In this approach, the tasks of estimating causal effects in the overlap and non-overlap regions are delegated to two distinct models, suited to the degree of data support in each region. Tree ensembles are used to non-parametrically estimate individual causal effects in the overlap region, where the data can speak for themselves. In the non-overlap region, where insufficient data support means reliance on model specification is necessary, individual causal effects are estimated by extrapolating trends from the overlap region via a spline model. The promising performance of our method is demonstrated in simulations. Finally, we utilize our method to perform a novel investigation of the causal effect of natural gas compressor station exposure on cancer outcomes. Code and data to implement the method and reproduce all simulations and analyses, is available on Github (\url{https://github.com/rachelnethery/overlap}).
\end{abstract}

\section{Introduction}
\label{s:intro}
\subsection{Natural Gas Compressor Stations and Cancer Mortality}
During the last several decades, the United States (US) has witnessed a sharp increase in the incidence of thyroid cancer, which now accounts for 1-1.5\% of all newly diagnosed cancer cases \citep{pellegriti2013worldwide}. Increased exposure of the population to radiation and carcinogenic environmental pollutants is blamed, in part, for this increase.

During the last several decades, US natural gas (NG) production has also increased rapidly. NG production and distribution systems have recently received attention as a potential source of human exposure to carcinogenic pollutants and endocrine-disrupting chemicals \citep{kassotis2016endocrine}. Recent epidemiological studies have found links between NG production and leukemia and between NG production and thyroid cancer \citep{finkel2016shale,mckenzie2017childhood}. The relationship between NG systems and thyroid cancer could be of particular interest due to their coincident rise.

Most previous studies of the health effects of NG systems have focused on associations between exposure to production sites (e.g., drilling wells) and health outcomes \citep{finkel2016shale,rasmussen2016association,mckenzie2017childhood}. In this study, we turn our attention instead to the potential health effects of NG distribution systems. Specifically, we aim to provide the first data-driven epidemiological investigation of the causal effects of proximity to NG compressor stations on thyroid cancer and leukemia mortality rates.

NG compressor stations are pumping stations located at 40-70 mile intervals along NG pipelines. They keep pressure in the pipelines so that NG flows in the desired direction \citep{messersmith2015}. The operations at compressor stations have raised health concerns for residents of nearby communities \citep{EHP2015}. In this paper, we exclude from consideration the health impacts of accidents at compressor stations and focus on the potentially harmful exposures to nearby communities resulting from the normal operations of compressor stations. Fugitive emissions, or unintended leaking of chemicals from the compressor station equipment, are known to occur but are not well-characterized. NG compressor stations also routinely conduct ``blowdowns'', in which pipelines and equipment are vented to reduce pressure \citep{EHP2015-2} and any chemicals present in the pipeline are reportedly released into the air in a 30-60 meter plume of gas \citep{EHP2015}. Little is known about the specific types of chemicals emitted.

While airborne emissions from compressor stations are regulated by the EPA under the Clean Air Act \citep{messersmith2015}, air quality studies in Pennsylvania and Texas have discovered harmful chemicals in excess of standards near NG compressor stations \citep{wolfeagle2009,padep2010}. These chemicals include methane, ethane, propane, and numerous benzene compounds. Benzene is a known carcinogen \citep{maltoni1989benzene,golding1999possible} and many of these compounds are known or suspected endocrine disruptors \citep{epaEDSP}.

Motivated by these findings, we present an investigation of the causal effects of compressor stations on county-level thyroid cancer and leukemia mortality rates. From a sample of 978 counties from the mid-western region of the US, we obtained their NG compressor stations exposure status, their thyroid cancer and leukemia mortality rates, and many suspected confounders of this relationship. While we would like to apply a classic non-parametric causal inference analysis rooted in the potential outcomes approach \citep{rubin_estimating_1974}, the data exhibit non-overlap, i.e., in some areas of the confounder space, there is little or no variability in the exposure status of the units. Due to this non-overlap, any attempt to adjust for confounding when estimating the population average causal effect must rely upon model-based extrapolation, because we have insufficient data to infer about missing potential outcomes in those regions of the confounder space. Thus non-parametric causal inference methods may yield unreliable results.

In this paper, we seek to make two methodological contributions to the causal inference literature, in the context of the potential outcomes framework. First, we introduce a flexible, data-driven definition of sample propensity score overlap and non-overlap regions. Second, we propose a novel approach to estimating population average causal effects in the presence of non-overlap. Using this approach, the sample is split into a region of overlap (RO) and a region of non-overlap (RN) and distinct models, appropriate for the amount of data support in each region, are developed and applied to estimate the causal effects in the two regions separately. We have found that the proposed approach leads to improved estimation of the population average causal effects compared to existing methods. Moreover, we apply this method to estimate the population average causal effect of compressor station exposure on thyroid cancer and leukemia mortality.

\subsection{Causal Inference Notation and Assumptions}

We first introduce notation that will be used throughout this article. For subject $i$, $(i=1,\cdots N)$, $Y_i^{obs}$ will denote the observed outcome (here it will be assumed to be a continuous random variable, in Section~\ref{ss:binaryout} we introduce analogous notation for the binary outcomes setting), $E_i$ will denote a binary treatment or exposure, and $\boldsymbol{X}_i$ will denote a vector of observed confounders. Under the stable unit treatment value assumption \citep{rubinrandomization1980}, potential outcomes $Y_i(1)$ and $Y_i(0)$, corresponding to the outcome that would be observed under scenarios $E_i=1$ and $E_i=0$, respectively, exist for each unit. Only one of these potential outcomes can be observed, such that $Y_i^{obs}=E_i Y_i(1)+(1-E_i) Y_i(0)$. We denote each unit's missing potential outcome as $Y_i^{mis}$, i.e., $Y_i^{mis}= (1-E_i) Y_i(1)+ E_i Y_i(0)$. An individual causal effect refers to the difference in potential outcomes for an individual, i.e., $\Delta_i=Y_i(1)-Y_i(0)$. The sample average causal effect is $\Delta_S=\frac{1}{N} \sum_{i=1}^N \Delta_i$, the population conditional average causal effect is $\Delta_{P|\boldsymbol{x}}=E[Y(1)-Y(0)|\boldsymbol{x}]$, and the population average causal effect is $\Delta_P=E_{\boldsymbol{X}}[E[Y(1)-Y(0)|\boldsymbol{X}]]$.

The identifiability of population level causal effects in observational studies typically relies upon the assumptions of (1) unconfoundedness and (2) positivity. Unconfoundedness implies that all confounders of the relationship between exposure and outcome are observed, i.e., $ E_i \independent (Y_i(1),Y_i(0)) |\boldsymbol{X}_i$. We assume throughout that unconfoundedness holds. Positivity, stated mathematically as $0<P(E_i=1|\boldsymbol{X}_i)<1$, is the assumption that each individual has positive probability of obtaining either exposure status. We also assume that positivity holds, i.e., that each individual in the population is eligible to receive either exposure status. A number of existing papers discuss positivity violations \citep{cole_constructing_2008,westreich_invited_2010,petersen_diagnosing_2012,damour2017overlap}.

We relax the assumption of overlap, closely related to positivity, which is required to non-parametrically estimate sample and population average causal effects. The term overlap refers to the overlap of the confounder distributions across the exposure groups. Non-overlap occurs when every unit in the population is eligible to receive either exposure, but, by chance, few or no units from one exposure group are observed in some confounder strata \citep{westreich_invited_2010}. Non-overlap can be a population level feature or a finite sample issue only. Here, we address problems arising from finite sample non-overlap, i.e., scenarios in which the population exhibits complete overlap but, in some areas of the confounder space, data are sparse for one or both exposure groups, leading to representative samples with non-overlap.

In the presence of non-overlap, sample and population average causal effect estimates generally suffer from bias and increased variance unless they are able to rely on the additional assumption of correct model specification \citep{king2005dangers,petersen_diagnosing_2012}. The overlap assumption can be evaluated by comparing the empirical distribution of the estimated propensity score, $\hat{\xi}_i=\hat{P}(E_i=1 \mid\boldsymbol{X})$, between the exposure groups \citep{rosenbaum_central_1983,austin_introduction_2011}, assuming the propensity score is well-estimated. The further assumption that non-overlap is a finite sample feature only is generally untestable but can sometimes be evaluated using subject-matter expertise. While the assumption of unconfoundedness becomes more plausible as the number of covariates grows, the likelihood of non-overlap increases \citep{cole_constructing_2008,damour2017overlap}; thus, non-overlap is an increasing problem in our era of high dimensional data.
  


\subsection{Existing Methods for Estimating Causal Effects in the Presence of Non-Overlap}

Methodological proposals for reducing the bias and variance of causal effect estimates in the presence of propensity score non-overlap are abundant in the causal inference literature \citep{cole_constructing_2008,crump_dealing_2009,petersen_diagnosing_2012,li_balancing_2016}; however, to our knowledge, all of the existing methods modify the estimand and its interpretation so that neither sample nor population average causal effect estimates can be obtained. In many medical and health research settings, such as evaluation of treatments, the aim of the research is to help clinicians choose between various forms of treatment for patients who are likely to adhere to any of the available treatments. In these contexts, convenience samples are common and modified estimands may be equally informative as or more informative than sample or population level estimands. However, in the environmental health applications like the one considered in this paper, study samples are often carefully selected to reflect a population of interest to policymakers. Here the primary aim is to estimate the burden of disease attributable to certain contaminants for the whole sample and/or underlying population in order to ultimately inform regulatory policies. Examples of such applications include the effect of power plant emissions on cardiovascular hospitalizations \citep{zigler2016hei}, the effect of unconventional natural gas extraction on childhood cancer incidence \citep{mckenzie2017childhood}, and the effect of gestational chemical exposures on birth outcomes \citep{ferguson2014environmental}.

The most commonly recommended approach for handling propensity score non-overlap is ``trimming'' or discarding observations in regions of poor data support \citep{ho2007matching,petersen_diagnosing_2012,gutman2015estimation}. Several papers provide guidance on how to determine which observations should be trimmed, most in the context of matching \citep{cochran1973controlling,lalonde1986evaluating,dehejia1999causal,ho2007matching,crump_dealing_2009}. An appealing feature of trimming is its flexibility-- it can be applied in conjunction with any causal estimation procedure. However, trimming allows only for the estimation of the average causal effect {\it in the trimmed sample}. Moreover, trimming changes the asymptotic properties of estimators in ways that are often overlooked \citep{yang2018asymptotic}.

Recent methodological developments in the context of weighting approaches to causal inference may provide more interpretable estimands than trimming. Li et al. (2\citeyear{li_balancing_2016}, 2\citeyear{li2018addressing}) introduce overlap weights, which weight each member of the sample proportional to its probability of inclusion in its counterfactual exposure group. The estimand corresponding to the overlap weights is what Li et al. call the ``average exposure effect for the overlap population'', and this overlap population, the sub-population that contains substantial proportions of both exposed and unexposed individuals, is of interest in many clinical and public health settings. \citet{zigler2017posterior} implement a similar overlap weighting approach in a Bayesian framework. 

In contrast to the existing literature, which emphasizes the removal or downweighting of data in regions of poor support, we propose a method that (1) minimizes model dependence where possible and (2) performs model-based extrapolation in a principled manner where necessary, yielding estimates of the population level estimand with small bias and appropriately large uncertainty. This method will be most valuable in environmental health and other applications where preserving population-level inference in spite of non-overlap is critical.

Our Bayesian modeling approach estimates individual causal effects ($\Delta_i$) in the RO and the RN separately. In the first stage of this procedure, a non-parametric Bayesian Additive Regression Tree (BART) \citep{chipman_bart_2010}, is fit to the data in the RO to estimate causal effects $\Delta_i$ for each observation in the RO, where data support is abundant. In the second stage, a spline (SPL) is fitted to the estimated $\Delta_i$ in the RO to capture trends in the causal effect surface. The SPL is used to extrapolate those trends to estimate $\Delta_i$ for observations in the RN, where insufficient data support requires reliance on model specifications/extrapolation. The data in the RN are excluded from all model fitting, so that the models are not influenced by data-sparse regions; however, after model fitting to data in the RO, the observed potential outcomes in the RN are employed as covariate values to aid in prediction of causal effects in the RN, so that we maximally leverage the sparse data in the RN as well. Because of the flexibility of both BART and SPL, we will show that our model captures non-linearities and causal effect heterogeneity.

In Section~\ref{s:methods}, we provide a data-driven definition of the RO and RN, and we introduce our method, called BART+SPL. Simulations in Section~\ref{s:sims} demonstrate how BART+SPL can yield improved population average causal effect estimates relative to existing methods and provide guidance in specifying tuning parameters. In Section~\ref{s:app}, BART+SPL is applied to estimate the effect of NG compressor station exposure on thyroid cancer and leukemia mortality rates. We conclude with a summary of our findings in Section~\ref{s:discuss}.

\section{Methods}
\label{s:methods}

\subsection{Definition of Overlap and Non-Overlap Regions}
\label{ss:overlap}

Although \citet{king2005dangers} proposed the use of the convex hull of the data to define overlap and non-overlap regions as early as 2005, this appears to be an unpopular criteria, likely due to its extreme conservatism. \citet{crump_dealing_2009} noted the absence of a systematic definition of the RO in the causal inference literature. While they offer a definition, their goal is to identify a region of the data that will produce a minimum variance average causal effect estimate rather than to provide a general definition of the region where overlap is observed. Their definition may be ideal in the context of trimming in conjunction with non-parametric causal estimators; however, it is unlikely to be appropriate in more general settings. BART itself has also been proposed as a method for identifying the RN \citep{hill2013assessing}. Because BART yields relatively large uncertainties for predicted values of units in regions of poor data support, \citet{hill2013assessing} suggest trimming observations with posterior uncertainty values greater than some threshold.

We intend to provide a more general characterization of the RO and RN here. We use the estimated propensity scores, $\hat{\xi}_i$, to define the RO, $O$, and the RN, $O^\perp$, in the sample. Throughout this section, we assume that the propensity score model has been correctly specified (or that the true propensity score is known, in which case $\xi_i$ can be substituted for $\hat{\xi}_i$ in the following); however, we demonstrate and discuss the performance of our method under propensity score model misspecification using simulations in Section~\ref{s:sims}. Let $\hat{\xi}_{(j)}$ denote the $j^{th}$ order statistic of the $\hat{\xi}_i$ and $P=\left[\hat{\xi}_{(1)},\hat{\xi}_{(N)}\right]$ be the subspace of $\left(0,1\right)$ over which the $\hat{\xi}_i$ are observed.

Our definition allows every point in $P$ to be assigned to either $O$ or $O^\perp$. The user must pre-specify two parameters, denoted $a$ and $b$, which are used to identify $O$. $O^\perp$ is then defined as the complement of $O$, relative to $P$. Consider any point $o \in P$. The idea behind our definition of overlap is that, if more than $b$ units from each exposure group have estimated propensity scores lying within some open interval of size $a$ covering $o$, then $o$ is included in the region of overlap. Thus, $a$ is an interval length, i.e., a portion of the range of the estimated propensity score, and $b$ is a portion of the sample size, representing the number of estimated propensity scores from each exposure group that must lie sufficiently close, i.e., within an interval of length $a$, to any given point in order for the point to be added to the RO. Framing this definition in a way that can be operationalized, it says that there is sufficient data support (overlap) at a point $o$ if, for each exposure group separately, we can form a set that includes (1) $o$ and (2) more than $b$ estimated propensity scores and lies entirely within an interval of size less than $a$, i.e., has range less than $a$.

We now introduce notation that will be used in the definition. Let $N_e$ denote the number of units in exposure group $e$ and $\hat{\xi}^e_{(i)}$ denote the $i^{th}$ propensity score order statistic in exposure group $e$. Using this notation, we propose the following definition for the region of overlap:
\[O=\left\lbrace o \in P: \text{for }e=0,1,\text{ range}(\left\lbrace o, \hat{\xi}^e_{(i)},...,\hat{\xi}^e_{(i+b)} \right\rbrace)<a \text{ for some } i=1,...,N_e-b  \right\rbrace\]
This formalizes the notion introduced above of finding a set of $o$ and more than $b$ propensity scores, $\left\lbrace o, \hat{\xi}^e_{(i)},...,\hat{\xi}^e_{(i+b)} \right\rbrace$, with range less than $a$ for each exposure group.

In comparison with previous definitions, our overlap definition provides a flexible, transparent, and data-driven approach to identifying regions of poor data support, with relatively easy-to-understand tuning parameters that give users the ability to decide what constitutes ``sufficient'' data support in the context of their own methods and application. Another contribution of this overlap definition to the wider literature is that it allows for regions of non-overlap in the interior of the propensity score distribution. In Figure 2 in Section A.3 of the Appendix, we provide an illustration of the types of non-overlap that can be captured by this definition. 

We define the RO and RN for BART+SPL using this definition. However, we note that BART+SPL is designed to handle exclusively non-overlap in the tails of the propensity score distribution. In Section~\ref{ss:choosea}, we provide guidance on how to specify $a$ and $b$ when applying BART+SPL. 

\subsection{Tree Ensembles for Causal Inference}
\label{ss:trees}
Regression trees are a class of non-parametric machine learning procedures, primarily used for prediction, in which a model is constructed by recursively partitioning the covariate space. Tree ensemble methods like BART \citep{chipman_bart_2010} gained popularity as a more stable generalization of regression trees. Appealing features of tree ensembles include (1) their internal construction of the model, eliminating the need to pre-specify the functional form of the association between the response and predictors and (2) their ability to capture complex non-linear associations and high order interactions \citep{strobl2009introduction}.

BART is highly regarded for its consistently strong performance under the ``default'' model specifications, reducing its dependence on subjective tuning and time consuming cross validation procedures. Letting $j$ index the $J$ trees in the ensemble ($j=1,...,J$), a BART is a sum of trees model of the form 
\begin{equation}
    Y=\sum_{j=1}^J g(\boldsymbol{X};\mathcal{T}_j,\mathcal{M}_j)+\epsilon,
\end{equation}
where $g$ is a function that sorts each unit into one of a set of $m_j$ terminal nodes, associated with mean parameters $\mathcal{M}_j=\{\mu_1,...,\mu_{m_j}\}$, based on a set of decision rules, $\mathcal{T}_j$. $\epsilon$ is a random error term that is typically assumed to be $N(0,\sigma^2)$ when the outcome is continuous. BART has also been extended to the binary outcome setting through the addition of the probit link function. BART's strong predictive performance has been reported in many contexts \citep{zhou2008extracting,he2009profiling,chipman_bart_2010,liu2010prediction,bonato2011bayesian,hill_bayesian_2011,huang2015predicting,liu2015ensemble,kindo2016multinomial,sparapani2016nonparametric}.

BART was first introduced as a tool for causal inference by \citet{hill_bayesian_2011}, who suggested fitting a BART, including the estimated propensity score as a covariate, and using it to predict missing potential outcomes. Despite BART's highly accurate potential outcome prediction in regions of the data with strong support, its predictions have been shown to sometimes contain greater bias than those of parametric and classic causal inference approaches in the presence of non-overlap \citep{hill_bayesian_2011,hill2013assessing}. Because BART relies on binary cuts of the observed predictors, it is unable to capture trends in the data and therefore extrapolates poorly.

\subsection{BART+SPL}
In this section, we describe BART+SPL, our proposed Bayesian approach for estimating causal effects in the presence of propensity score non-overlap. The first stage of the procedure, which we call the imputation phase, utilizes a BART to impute the missing potential outcomes and estimate individual causal effects in the RO. In the second stage, which we call the smoothing stage, a spline is fit to the BART-estimated individual causal effects in the RO and is invoked to extrapolate the causal effect trends to the individuals in the RN, leveraging the information from the observed potential outcomes for observations in the RN. Our approach for continuous outcomes is described in Sections~\ref{sss:imputation} and ~\ref{sss:smoothing} in the context of a single iteration of a Bayesian MCMC sampler for the sake of clarity, and in Section~\ref{sss:estimation}, we explain how the draws from the sampler can be invoked to estimate causal effects and associated uncertainties. This model can be implemented through the MCMC procedure described in Section A.1 of the Appendix. In Section~\ref{ss:binaryout}, we introduce an extension to binary outcomes.

\subsubsection{Imputation Stage}
\label{sss:imputation}
In the first stage of BART+SPL, we adopt the common practice of treating the unobserved potential outcome for each individual as missing data, and we construct a BART model to impute these missing values for individuals in the RO. We introduce the subscripts $q$ and $r$ to index data from subjects in the RO and RN, respectively, e.g., $Y_{q}^{obs}$ is the observed outcome of individual $q$ in the RO and $Y_{r}^{mis}$ is the missing potential outcome of individual $r$ in the RN $(q=1,...,Q;r=1,...,R)$. Subscript $O$ and subscript $O^\perp$ refer to vectors/matrices of the values of all individuals in $O$ and $O^\perp$, respectively, e.g., $\mathbf{Y}_{O}^{mis}=\left[Y_1^{mis},...,Y_Q^{mis}\right]'$ and $\mathbf{Y}_{O^\perp}^{mis}=\left[Y_1^{mis},...,Y_R^{mis}\right]'$.

In this stage, all of our modeling efforts are focused on the data in the RO. $Y_{q}^{mis}$ is first imputed using a BART model of the form
\[
    Y_{q}^{obs}=\sum_{j=1}^J g(E_{q},\hat{\xi}_{q},\boldsymbol{X}_{q};\mathcal{T}_j,\mathcal{M}_j)+\epsilon_{q},
\]
where $\epsilon_{q}\sim N(0,\sigma_B^2)$. To do so, the Bayesian backfitting algorithm of \citet{chipman_bart_2010} is utilized to collect a sample from the posterior distribution of $\theta=\{\sigma_B^2,\mathcal{T}_j,\mathcal{M}_j; j=1,...,J\}$, $p(\theta|\mathbf{Y}_{O}^{obs})$. An imputed value of $\mathbf{Y}_{O}^{mis}$, denoted $\tilde{\mathbf{Y}}_{O}^{mis}$, is obtained by sampling from its posterior predictive distribution (ppd), $p(\mathbf{Y}_{O}^{mis}|\mathbf{Y}_{O}^{obs})=\int p(\mathbf{Y}_{O}^{mis}|\mathbf{Y}_{O}^{obs},\theta)p(\theta|\mathbf{Y}_{O}^{obs})d\theta$.  $\mathbf{Y}_{O}^{obs}$ and $\tilde{\mathbf{Y}}_{O}^{mis}$ are used to construct a sample of the individual causal effects in $O$, $\tilde{\mathbf{\Delta}}_{O}$.

\subsubsection{Smoothing Stage}
\label{sss:smoothing}
In the second stage, a smoothing model is fit to the BART-estimated individual causal effects in the RO, and the model is employed to estimate the individual causal effects in the RN by extrapolating the trends identified in the RO. With this approach, we impose the assumption that any trends in the individual causal effects (as a function of the propensity score and/or the covariates) identified in the RO can be extended into the RN. By modeling the causal effect surface in this stage rather than the separate potential outcome surfaces, we take advantage of the potentially increased smoothness of the causal effects that may occur in practice. Through tuning, we ensure that the variance in the RN is inflated to reflect the high uncertainty in the region, leading to mildly conservative confidence regions in most realistic scenarios.

Assume for now that the RN includes only exposed individuals so that $E_r=1$ for all $r$. Define $Y_{q}^*(1)=Y_{q}^{obs}$ if $E_{q}=1$ and $Y_{q}^*(1)=\tilde{Y}_{q}^{mis}$ otherwise (remember $\tilde{Y}_{q}^{mis}$ denotes the BART-imputed missing potential outcome for $q$). Thus, $Y_{q}^*(1)$ is the observed or imputed potential outcome corresponding to $E=1$ for each unit in the RO. Let $rcs(z)$ denote a restricted cubic spline basis for $z$. Employing the imputed values obtained in the previous stage, we construct the following smoothing stage model:
\[
\tilde{\Delta}_{q}=\boldsymbol{W' \beta}+\epsilon_{q}, \quad \boldsymbol{W}=\left[\begin{array}{c}
rcs(\hat{\xi}_{q})\\
rcs(Y_{q}^*(1))\\
\boldsymbol{X}_{q}
\end{array}\right],
\]
where $\epsilon_{q}\sim N(0,\sigma_S^2+I(\hat{\xi}_{q}\in O^\perp)\tau_{q})$. $\sigma_S^2$ is the residual variance for all units in the RO and $\tau_{q}$ is an added variance component only applied to units in the RN. The purpose of $\tau_{q}$ is to inflate the variance of units with an estimated propensity score in the RN, to adequately reflect the higher uncertainty in regions of little data support. In this model, $\tau_{q}$ is clearly unidentifiable, as the model is fit using exclusively data in the RO, and it will only come into play when invoking the ppd to predict in the RN. Here, we choose to treat it as a tuning parameter, and below we describe our recommended tuning parameter specification. We recommend restricted cubic splines in the smoothing model, because they generally demonstrated superior performance when applied to simulated data, however other spline choices may provide improved performance under some conditions. We also recommend excluding a small portion of the data at the tails of the RO (i.e., tails of the propensity score in the RO) from spline model fitting, either through the use of boundary knots or by omitting these data from the model, because BART's predictions in the tails of variables can be unstable and can negatively influence the SPL's performance.

We collect a posterior sample of the spline parameters, $\psi=\{\boldsymbol{\beta},\sigma_S^2\}$. Recall the motivation for the smoothing stage is to use the trends from the RO to predict the individual causal effects in the RN. Thus, the sampled $\psi$ and the covariate values for individuals in the RN are summoned to obtain a sample from the ppd of $\mathbf{\Delta}_{O^\perp}$, $p(\mathbf{\Delta}_{O^\perp}|\mathbf{\Delta}_{O})=\int p(\mathbf{\Delta}_{O^\perp}|\mathbf{\Delta}_{O},\psi)p(\psi|\mathbf{\Delta}_{O})d\psi$. The sample is denoted $\tilde{\mathbf{\Delta}}_{O^\perp}$.

Note that including $Y_{q}^*(1)$ as a predictor permits the model to capture the relationship between the causal effects and $Y(1)$, the potential outcome observed for all units in the RN. This allows the observed potential outcomes in the RN to aid in the extrapolation, so that this information is not wasted. In the case that both exposed and unexposed units fall in the RN, we define $Y_{q}^*(0)$ analogously to $Y_{q}^*(1)$ and construct a second model, identical to the one above except replacing $rcs(Y_{q}^*(1))$ with $rcs(Y_{q}^*(0))$. The ppd from the first model is then used to predict individual causal effects for exposed units in the RN and the ppd from the second model is used for unexposed units.

Our recommended specification of $\tau_{q}$ is motivated by the aim to have (1) the variance of individual causal effects increase monotonically as the observation's distance from the RO (i.e., region of strong data support) increases and (2) the increase in variance be in proportion to the scale of the data. Thus, the suggested tuning parameter specification is $\tau_{q}=(10d_{q})t_O$, where $t_O=\text{range}(\tilde{\mathbf{\Delta}}_{O})$ and $d_{q}$ is the distance from the observation's propensity score to the nearest propensity score in the RO. The effect of this tuning parameter is that, for every .1 unit further we go into the RN, the variance of the individual causal effects increases by the range of the causal effects in the RO. While it could lead to somewhat conservative credible interval coverage in ``simple'' situations (e.g., when the trends in the RN are easily predicted from the trends in the RO), we have found in simulations that this choice of tuning parameter consistently provides both reasonably-sized credible intervals and acceptable coverage.

\subsubsection{Estimation and Uncertainty Quantification}
\label{sss:estimation}

We can iterate the two stages described above $M$ times to obtain $\left\lbrace \mathbf{\Delta}^{(1)}_{O},...,\mathbf{\Delta}^{(M)}_{O} \right\rbrace$ from the imputation stage and $\left\lbrace \mathbf{\Delta}^{(1)}_{O^\perp},...,\mathbf{\Delta}^{(M)}_{O^\perp}\right\rbrace$ from the smoothing stage (note that we have traded the tilde notation from above for the $(m)$ notation to differentiate the samples from the $M$ iterations). By iterating between the two stages, we are able to account for the uncertainty in the estimation of $\mathbf{\Delta}_O$ from the first stage and pass it on to the second stage, where $\mathbf{\Delta}_O$ is used as the outcome. Thus, the uncertainty in the estimate of $\mathbf{\Delta}_{O^\perp}$ reflects the uncertainty both from stage one and stage two.

For units in the RO and the RN, individual causal are estimated as $\hat{\Delta}_{q}=\frac{1}{M}\sum_{m=1}^M \Delta_{q}^{(m)}$ and $\hat{\Delta}_{r}=\frac{1}{M}\sum_{m=1}^M \Delta_{r}^{(m)}$, respectively, i.e., the posterior mean over the $M$ samples. Credible intervals for the individual causal effects can be obtained by extracting the appropriate percentiles from these $M$ samples. Samples of $\Delta_S$ are produced by  $\Delta^{(m)}_S=\frac{1}{N}(\sum_{q=1}^Q \Delta^{(m)}_{q} + \sum_{r=1}^R \Delta^{(m)}_{r})$ for $m=1,...,M$, and $\hat{\Delta}_S=\frac{1}{M} \sum_{m=1}^M \Delta^{(m)}_S$. As above, percentiles of the $M$ samples provide credible interval for $\Delta_S$.

In order to estimate $\Delta_P$, an additional integration over the predictors is required. \citet{wang2015accounting} discuss the necessity of such an integration step when estimating population average causal effects with models that permit non-linearity and/or heterogeneity, and they propose the application of the Bayesian bootstrap to execute it. We adopt the same approach here. For each sample of the individual causal effects, $\left\lbrace \mathbf{\Delta}^{(m)}_{O}, \mathbf{\Delta}^{(m)}_{O^\perp} \right\rbrace$, the Bayesian bootstrap is performed on it $B$ times (where $B$ is a large constant) and the average of each bootstrap sample taken to obtain $B$ draws from the posterior distribution of the population average causal effect. We randomly select one of these samples and call it $\Delta^{(m)}_P$, so that in the end we have collected $\left\lbrace \Delta^{(1)}_{P},...,\Delta^{(M)}_{P} \right\rbrace$. The population average causal effect is estimated as $\hat{\Delta}_P=\frac{1}{M} \sum_{m=1}^M \Delta^{(m)}_P$ and the credible interval formed using percentiles.

\subsection{BART+SPL with Binary Outcomes}
\label{ss:binaryout}

By invoking BART probit \citep{chipman_bart_2010} in the imputation stage and utilizing a simple arcsine transformation in the smoothing stage, we can straightforwardly extend BART+SPL to the binary outcomes setting. While most of our notation will remain the same for binary outcomes, we note a few changes. Individual causal effects are traditionally defined as the difference in each individual's potential outcomes, as above; however, in the binary outcomes setting, estimating those differences, which can only take on values -1, 0 and 1, may be challenging and may sacrifice information. Because, in our approach, we are treating the potential outcomes as random variables, it is reasonable and desirable to instead define the individual causal effects as differences in some features of the distributions of the potential outcomes, although doing so requires a slight abuse of traditional terminology/notation. For binary outcomes, we define the individual causal effects as $\Delta_i=P(Y_i(1)=1)-P(Y_i(0)=1)$ and the estimands as $\Delta_S=\frac{1}{N} \sum_{i=1}^N \Delta_i$, $\Delta_{P|\boldsymbol{x}}=P(Y(1)=1|\boldsymbol{x})-P(Y(0)=1|\boldsymbol{x})$, and $\Delta_P=E_{\boldsymbol{X}}[P(Y(1)=1|\boldsymbol{X})-P(Y(0)=1|\boldsymbol{X})]$. Here we fit a BART probit to estimate individual causal effects in $O$, fit a spline model to the arcsine transform of these estimates (which are bounded between -1 and 1), and use the spline to estimate individual causal effects for units in $O^\perp$. While we provide below explicit forms for the imputation and smoothing models in the binary setting, we refer the reader back to the previous section for the full sampling procedure details, which follow analogously to the continuous outcomes setting.

In the imputation stage, the BART probit model fit to the RO data has the following form:
\[
P(Y_{q}^{obs}=1)=\Phi(\sum_{j=1}^J g(E_{q},\hat{\xi}_{q},\boldsymbol{X}_{q};\mathcal{T}_j,\mathcal{M}_j)),
\]
where $\Phi()$ is the standard Normal cumulative distribution function. With this model, posterior samples $\tilde{P}(Y_{q}^{obs}=1)$ and $\tilde{P}(Y_{q}^{mis}=1)$ can be drawn and used to form a posterior sample of the individual causal effect, $\tilde{\Delta}_{q}$.

For the smoothing stage, as above, assume without loss of generality that all the units in $O^\perp$ all have $E_r=1$. Define $Y_{q}^*(1)=Y_{q}^{obs}$ if $E_{q}=1$ and $Y_{q}^*(1)=I(\tilde{P}(Y_{q}^{mis}=1)>0.5)$ otherwise. Then the smoothing model is
\[
\text{arcsine}(\tilde{\Delta}_{q})=\boldsymbol{W' \beta}+\epsilon_{q}, \quad \boldsymbol{W}=\left[\begin{array}{c}
rcs(\hat{\xi}_{q})\\
Y_{q}^*(1)\\
\boldsymbol{X}_{q}
\end{array}\right],
\]
where $\epsilon_{q}\sim N(0,\sigma_S^2)$. Note that, unlike in the continuous case, no tuning parameter is included in the variance, as simulations indicated it was not needed to obtain reasonable coverage in the binary setting. Individual causal effects on the arcsine scale for units in $O^\perp$ can be obtained from the posterior predictive distribution and back-transformed to the desired scale. As described in the previous section, a second analogous smoothing model can be fit if $O^\perp$ contains units from both the exposed and unexposed groups. Average causal effect estimation and uncertainty quantification proceed identically to the continuous case.

\section{Simulations}
\label{s:sims}

In this section, we conduct simulation studies to evaluate the performance of BART+SPL relative to existing methods in the presence of non-overlap and to provide guidance on how to specify parameters $a$ and $b$ in the non-overlap definition to obtain optimal performance. In Section~\ref{ss:performance}, we simulate data with a small number of confounders and varying degrees of non-overlap, and we compare BART+SPL's population average causal effect estimation performance to that of a standard BART and of an existing spline-based method for causal inference. In Section ~\ref{ss:hdperformance}, we generate data with non-overlap and high dimensional covariates and compare the performance of BART+SPL and the spline-based method. Finally, data with varying amounts of non-overlap are generated and BART+SPL is implemented with various specifications of $a$ and $b$ to provide insight on the optimal choices in Section~\ref{ss:choosea}. R code \citep{R2016} to implement BART+SPL and to reproduce all simulations is available on Github at \url{https://github.com/rachelnethery/overlap}.

\subsection{Performance of BART+SPL Relative to Existing Methods}
\label{ss:performance}

We purposely simulate data under a challenging situation of: a) propensity score non-overlap; b) non-linearity of the potential outcomes in the propensity score; and c) heterogeneous causal effects. We wish to evaluate the relative performance of our method when utilizing a true propensity score and when utilizing a misspecified propensity score estimate. We first discuss the simulation structure when utilizing the true propensity score. We let $N=500$ and assign half of the subjects to $E=1$. We generate two confounders that are highly associated with the exposure $(E)$, one binary $(X_1: X_1|E=1\sim Bernoulli(.5), X_1|E=0\sim Bernoulli(.4))$ and one continuous $(X_2: X_2|E=1\sim N(2+c,\sqrt{1.25+0.1c}), X_2|E=0\sim N(1,1))$. Given these specifications, the true propensity scores can easily be calculated using Bayes Rule. The potential outcomes are constructed as $Y_i(1)=-3(1+\exp(-(10(X_{2i}-1)))^{-1}+0.25X_{1i}-X_{1i}X_{2i}$ and $Y_i(0)=-1.5X_{2i}$. We label the simulations with the true propensity score as 3.1A.

For the simulations with a misspecified propensity score estimate, we again let $N=500$ and assign half of the subjects to $E=1$. We generate a binary confounder $(X_1: X_1|E=1\sim Bernoulli(.5), X_1|E=0\sim Bernoulli(.4))$ and a continuous confounder $(X_2: X_2|E=1\sim N(2+c,4), X_2|E=0\sim N(1,1))$. The potential outcomes are $Y_i(1)=3(1+\exp(-(10(X_{2i}-1)))^{-1}+0.25X_{1i}-0.1X_{1i}X_{2i}+0.5$ and $Y_i(0)=0.2X_{2i}+0.1X_{2i}^2+1$.
As is common in the literature, we use a simple logistic regression model of the form $\text{logit}(P(E_i=1))=\beta_0+\beta_1X_{1i}+\beta_2X_{2i}$ to estimate the propensity scores. This model is clearly misspecified, because, for example, the true relationship between $E$ and $X_2$ is not linear. In practice, when the form of the propensity score model is unknown, we encourage the use of flexible models for estimation \citep{westreich2010propensity}, such as BART, neural networks, or support vector machines, in order to reduce the chance of propensity score model misspecification. The use of BART for propensity score estimation is demonstrated in the application to real data in Section~\ref{s:app}. Various flexible propensity score estimation methods could be tested and the method that achieves the best covariate balance selected. We label the simulations with the misspecified propensity score as 3.1B.

We have selected these simulation structures so that, for a single value of $c$, the type and degree of propensity score non-overlap in 3.1A and 3.1B should be similar. Both 3.1A and 3.1B produce data sets with lack of overlap in the right tail of the propensity score distribution (i.e., individuals from the unexposed group are unobserved or very sparse), and with varying degrees of non-overlap controlled by $c$. Our simulations are designed to produce non-overlap in the right tail of the propensity score distribution and our motivation is to demonstrate how our method performs in the presence of different features in this RN. Thus, simulated datasets are utilized in the results below only if any intervals of non-overlap outside the right tail contain 10 observations or fewer (cumulatively), and, in these datasets, the intervals of non-overlap outside the right tail are ignored (i.e., treated as part of the RO). In this way, we ensure that the results solely reflect how the tested methods respond to the features of the intended RN.

We consider three separate simulated scenarios, i.e. three different specifications of $c$, within 3.1A and 3.1B. We let $c=0$ (simulations 3.1A-i and 3.1B-i), $c=0.35$ (simulations 3.1A-ii and 3.1B-ii), and $c=0.7$ (simulations 3.1A-iii and 3.1B-iii).  Example datasets from each are illustrated in Figure 3 in Section A.3 of the Appendix. With $c=0$, the RN is quite small, and the trend in the individual causal effects in the RN is mildly non-linear. With $c=0.35$, the RN is somewhat larger and the trends exhibited by the individual causal effects in the RN are moderately non-linear. With $c=0.7$, a substantial portion of the sample lies in the RN and the causal effects in the RN are highly non-linear. We use our definition of overlap with $a=.1$ and $b=7$ to define the RO and RN for each simulated dataset.

We implement BART+SPL on 1,000 simulated datasets under each condition. \citet{gutman2015estimation} recommended a spline-based multiple imputation approach for estimating average causal effects. They also suggest trimming in conjunction with their method for samples suffering from non-overlap. We compared the performance of BART+SPL versus Gutman and Rubin's method both with and without trimming (T-GR and U-GR, respectively) and also BART with and without trimming (T-BART and U-BART, respectively). Detailed results of the untrimmed analyses appear in Table~\ref{tab:results1} and the distributions of the average causal effect estimates from the trimmed and untrimmed analyses can be compared in Figure 4 in Section A.3 of the Appendix.

The simulation results demonstrate the dominant performance of BART+SPL compared to U-BART and U-GR under a wide range of challenging conditions. However, in extreme scenarios with unpredictable trends and large portions of the sample in the RN, even the performance of BART+SPL may deteriorate, as demonstrated by simulation 3.1A-iii, where BART+SPL gives high percent bias, and simulation 3.1B-iii, where BART+SPL gives high percent bias and poor coverage. Nonetheless, BART+SPL's performance still exceeds that of its competitors. For both BART and GR, the trimmed estimates, which are no longer estimators of the population level causal effects, are further from the true population average causal effects than the untrimmed estimates.

\begin{table}[ht]
\centering
\caption{Absolute (Abs) bias, 95\% credible interval coverage and mean square error (MSE) in estimation of the population average causal effects in simulations from Section 3.1.}
\begin{tabular}{crrrrr}
  \hline
Simulation Setting & Method & Abs Bias & Abs Bias (\%) & Coverage & MSE \\ 
  \hline
 \multirow{3}{*}{3.1A-i} & U-GR & 0.12 & 46.46 & 0.33 & 1.25 \\ 
 & U-BART & 0.03 & 10.72 & 0.99 & 0.07 \\ 
 & BART+SPL & 0.01 & 5.63 & 1.00 & 0.05 \\  \hline
 \multirow{3}{*}{3.1A-ii} & U-GR & 0.17 & 97.31 & 0.23 & 1.69 \\ 
 & U-BART & 0.05 & 31.95 & 0.89 & 0.12 \\ 
 & BART+SPL & 0.02 & 12.58 & 1.00 & 0.09 \\  \hline
  \multirow{3}{*}{3.1A-iii} & U-GR & 0.23 & 724.78 & 0.19 & 2.34 \\ 
 & U-BART & 0.08 & 427.55 & 0.71 & 0.22 \\ 
 & BART+SPL & 0.03 & 100.80 & 1.00 & 0.14 \\ \hline
 \multirow{3}{*}{3.1B-i} & U-GR & 0.26 & 50.42 & 0.00 & 0.64 \\
  & U-BART & 0.13 & 25.47 & 0.12 & 0.33 \\
  & BART+SPL & 0.11 & 21.64 & 0.62 & 0.27 \\ \hline
  \multirow{3}{*}{3.1B-ii} & U-GR & 0.32 & 66.58 & 0.00 & 0.73 \\
  & U-BART & 0.18 & 36.73 & 0.04 & 0.48 \\
  & BART+SPL & 0.15 & 30.95 & 0.55 & 0.36 \\ \hline
  \multirow{3}{*}{3.1B-iii} & U-GR & 0.41 & 94.79 & 0.00 & 0.89 \\
  & U-BART & 0.24 & 55.00 & 0.01 & 0.68 \\
  & BART+SPL & 0.21 & 48.03 & 0.35 & 0.49 \\
  \hline
\end{tabular}
\label{tab:results1}
\end{table}

We also conducted a simulation study to evaluate the performance of BART+SPL for binary outcomes. The simulated data structures utilized were similar to those described above. The data and the results are described in Section A.2 of the Appendix. BART+SPL performed similar to or better than the competing methods (U-GR and U-BART) in each of our simulations with $N=500$. However, even without non-overlap, BART probit can fail to provide improvements over parametric methods when sample sizes are small to moderate, and thus we recommend that BART+SPL for binary outcomes only be applied to large datasets (i.e., $N \geq 500$).

\subsection{BART+SPL with High Dimensional Covariates}
\label{ss:hdperformance}

One of the most widely-recognized limitations of BART, which was noted by its developers \citep{chipman_bart_2010}, is its poor performance when the number of predictors, $p$, is large. The decline in performance is most significant when many irrelevant predictors (i.e., predictors unrelated to the outcome) are included. Thus, in this section, we seek to examine whether and how BART+SPL should be applied in settings where the number of potential confounders is large. Although BART has been extended to permit sparsity in the $p>N$ setting \citep{linero_bayesian_2016}, we do not consider the $p>N$ case here.

For these simulations, we let $N=500$ and assign half of the sample to $E=1$. We then generate 10 confounders, 5 binary and 5 continuous. The binary confounders have distribution $X_1|E=1,...,X_5|E=1\sim Bernoulli(.45)$, $X_1|E=0,...,X_5|E=0\sim Bernoulli(.4)$, and the continuous confounders have distribution $X_6|E=1,...,X_{10}|E=1\sim N(2,4)$, $X_6|E=0,...,X_{10}|E=0\sim N(1.3,1)$. We consider the following three scenarios: only these 10 confounders are present (simulation 3.2A), these 10 confounders as well as 25 randomly generated ``potential confounders'' are present (simulation 3.2B), and these 10 confounders as well as 50 randomly generated ``potential confounders'' are present (simulation 3.2C). Of course, in real applications, we often do not know a priori which of the potential confounders are true confounders, hence we include them all in the modeling. A propensity score is formed using predicted probabilities from the logistic regression $\text{logit}(P(E_i=1))=\beta_0+\boldsymbol{Z}_i\boldsymbol{\beta}$, where $\boldsymbol{Z}_i$ is a vector of the true and potential confounders. The potential outcomes are generated so that they exhibit non-linear trends in the estimated propensity score-- $Y_i(0)=.5(X_{1i}+X_{2i}+X_{3i}+X_{4i}+X_{5i})+15(1+exp(-8X_{6i}+1))^{-1}+X_{7i}+X_{8i}+X_{9i}+X_{10i}-5$ and $Y_i(1)=X_{1i}+X_{2i}+X_{3i}+X_{4i}+X_{5i}-.5(X_{6i}+X_{7i}+X_{8i}+X_{9i}+X_{10i})$.

The features of these data are illustrated in Figure 6 in Section A.3 of the Appendix. The simulations are designed to have a large RN in the right tail of the estimated propensity score, with moderate non-linearity in the causal effect in the RN. The RO and RN are defined using tuning parameters $a=.1$ and $b=7$.

We simulate 1,000 datasets from each of the three scenarios described above. We apply both BART+SPL and the untrimmed Gutman and Rubin spline method (GR) to each dataset. Results are provided in Table~\ref{tab:hd}.

\begin{table}[ht]
\centering
\caption{Absolute (Abs) bias, 95\% credible interval coverage and mean square error (MSE) in estimation of the population average causal effects in simulations from Section 3.2.}
\begin{tabular}{crrrrr}
  \hline
Simulation Setting & Method & Abs Bias & Abs Bias (\%) & Coverage & MSE \\ 
  \hline
\multirow{2}{*}{3.2A}& GR &  0.9801 & 5.6382 & 0.0320 & 13.1447 \\ 
  & BART+SPL &  0.5608 & 3.2198 & 0.9850 & 8.9744 \\   \hline
\multirow{2}{*}{3.2B}& GR & 0.5627 & 3.2363 & 0.4000 & 14.3195 \\ 
  & BART+SPL & 0.5574 & 3.2012 & 0.9740 & 8.4942 \\  \hline
\multirow{2}{*}{3.2C}& GR &  0.3640 & 2.0958 & 0.7280 & 16.0908 \\ 
  & BART+SPL & 0.6368 & 3.6591 & 0.9570 & 9.6857 \\ 
   \hline
\end{tabular}
\label{tab:hd}
\end{table}

These results reflect BART's struggle in the presence of irrelevant predictors. When only the 10 true confounders are included in the modeling, BART+SPL outperforms GR and demonstrates similar performance as in Section~\ref{ss:performance}. However, when irrelevant predictors are introduced, GR's bias decreases while BART+SPL's remains constant or increases. With 50 irrelevant predictors, GR's bias is substantially lower than BART+SPL's (although, notably, its coverage and MSE remain inferior). These results should serve as a warning that BART+SPL is only likely to improve on existing methods in settings where the set of true confounders can be posited a priori with some confidence.

\subsection{Guidelines for Defining the RN}
\label{ss:choosea}

The simulation results in this section are intended to provide guidance on both the degree of non-overlap that threatens BART's performance and the degree of non-overlap that threatens BART+SPL's performance. They also suggest appropriate default specifications of tuning parameters $a$ and $b$ in the non-overlap definition. To impose strict control on the size of the RO and RN, in these simulations we utilize a single confounder rather than a propensity score. Based on the above simulations, we expect the performance of BART+SPL to be comparable with varying numbers of confounders, as long as few irrelevant covariates are included. Thus, the insights gained from these simulations are likely to scale well to scenarios with larger confounder sets.

We let $N=500$ and assign half of the sample to $E=1$. We generate the confounder as $X|E=1\sim N(2.5,4),\text{ } X|E=0\sim N(v,w)$, where $v$ and $w$ control the degree of non-overlap. Unlike in the previous simulations, in these we generate a single, fixed instance of the confounder and simply add random noise to (a function of) it to create the potential outcomes for each simulation. We consider two potential outcome scenarios, one of which produces data that are relatively simple to model (with BART) while the other produces data that are challenging to model. The former, which we label simulation 3.3A, is created by assigning $Y(0)=1.5+\frac{X+(X^2/2!)}{20}+\text{N}(0,0.06)$ and $Y(1)=\frac{1}{1+e^{-(X-1)}}+\text{N}(0,0.06)$ and the latter, which we call simulation 3.3B, by $Y(0)=1.5+\frac{X+(X^2/2!)+(X^3/3!)}{20}+\text{N}(0,0.06)$ and $Y(1)=\frac{1}{1+e^{-(X-1)}}+\text{N}(0,0.06)$.

In both simulation 3.3A and 3.3B, we achieve different degrees of non-overlap, primarily non-overlap in the right tail of the confounder, by manipulating $v$ and $w$. In order from least to most non-overlap, we consider $\left\lbrace v=1.4,w=1.96\right\rbrace$, $\left\lbrace v=0.75,w=1.44\right\rbrace$, and $\left\lbrace v=0,w=1\right\rbrace$. Moreover, in each scenario, we test the following three specifications of $\left\lbrace a,b\right\rbrace$ in the overlap definition, in order from most to least conservative: $\left\lbrace a=0.05*(range(X)),b=10 \right\rbrace$, $\left\lbrace a=0.1*(range(X)),b=10 \right\rbrace$, and $\left\lbrace a=0.15*(range(X)),b=3 \right\rbrace$. Considering each combination of $\left\lbrace v,w\right\rbrace$ and $\left\lbrace a,b\right\rbrace$ leads to 9 different settings for each of simulation 3.3A and 3.3B, for a total of 18 simulation settings. The proportion of the sample falling into the RN, denoted $\pi$ in these simulations ranges from $\pi=2$\% to $\pi=34$\%. Of course, the impact of non-overlap on average causal effect estimates depends not only on the proportion of the sample falling in the RN but also likely on the extremity of the observations in the RN relative to the RO. In our simulations, as $\pi$ increases, the average distance between observations in the RO and the RN also increases. An example dataset from both simulation 3.3A and simulation 3.3B is presented in Figure 7 in Section A.3 of the Appendix.

\begin{table}[h!]
\centering
\caption{Absolute (Abs) bias, 95\% credible interval coverage, and mean square error (MSE) in estimation of the population average causal effects using BART+SPL and BART applied to simulation 3.3A. RO-1 refers to simulations with the RO defined as $a=0.05*(range(X)),b=10$, RO-2 refers to simulations with the RO defined as $a=0.1*(range(X)),b=10$, and RO-3 refers to simulations with the RO defined as $a=0.15*(range(X)),b=3$.}
\begin{tabular}{rrrrrrr}
  \hline
$v,w$ & Method & $\pi$ & Abs Bias & Abs Bias (\%) & Coverage & MSE \\ 
  \hline
\multirow{4}{*}{$v=1.4,w=1.96$} & BART+SPL, RO-1 & 17 & 0.03 & 2.51 & 1.00 & 0.08 \\ 
&  BART+SPL, RO-2 & 8 & 0.03 & 2.23 & 1.00 & 0.07 \\ 
& BART+SPL, RO-3 & 2 & 0.03 & 2.43 & 1.00 & 0.08 \\
& BART & & 0.03 & 2.98 & 0.81 & 0.09 \\ \hline
\multirow{4}{*}{$v=0.75,w=1.44$}  & BART+SPL, RO-1 & 24 & 0.05 & 4.14 & 1.00 & 0.09 \\
&  BART+SPL, RO-2 & 14 & 0.04 & 3.44 & 1.00 & 0.08 \\
& BART+SPL, RO-3 & 6 & 0.05 & 4.06 & 1.00 & 0.09 \\
& BART & & 0.06 & 5.50 & 0.47 & 0.11 \\ \hline
\multirow{4}{*}{$v=0,w=1$}  & BART+SPL, RO-1 & 34 & 0.08 & 6.65 & 1.00 & 0.11 \\
&  BART+SPL, RO-2 & 21 & 0.05 & 4.60 & 1.00 & 0.09 \\
& BART+SPL, RO-3 & 11 & 0.06 & 4.71 & 1.00 & 0.09 \\
& BART & & 0.07 & 6.28 & 0.59 & 0.11 \\
   \hline
\end{tabular}
\label{tab:chooseaA}
\end{table}

We apply a standard BART (ignoring the non-overlap) and BART+SPL to $1,000$ simulated datasets under each of the 18 conditions. Table~\ref{tab:chooseaA} and Table~\ref{tab:chooseaB} contain the results for simulations 3.3A and 3.3B, respectively. While BART+SPL nearly always performs better in terms of each metric than BART, the most notable difference in the BART+SPL and BART results is the difference in coverage probabilities, with BART+SPL consistently obtaining conservative coverage and BART's coverage deteriorating as the degree of non-overlap increases. Even when only 2\% of the data falls into the RN, BART's coverage is unreliable. Thus, BART could provide misleading inference even with small amounts of non-overlap.

BART+SPL's coverage is conservative, but reliable, in all the simulations assessed, however its bias tends to increase as the degree of non-overlap increases. Thus, it appears that, if some bias in the point estimate can be tolerated, BART+SPL can be expected to provide conservative inference in (non-pathological) scenarios with over 25\% of the data in the RO. However, based on the observation that BART+SPL's bias is greater than 5\% in simulation 3.3B under each overlap definition with $\left\lbrace v=0.75,w=1.44 \right\rbrace$ and $\left\lbrace v=0,w=1 \right\rbrace$, a more conservative option might be to sacrifice the population-level estimand and performed a trimmed or weighted analysis when more than 15\% of the data falls in the RN.

Finally, we note that these simulations suggest that BART+SPL is quite robust to the specification of $a$ and $b$, as we see relatively small discrepancies in bias and coverage. However, some of the results of 3.3B indicate that one should avoid defining the RO too conservatively, as discarding too much information can lead to modest increases in bias. The moderate choice of $a=0.1*range(X)$ and $b=10$ provides the best results in most of the simulations, and we, therefore, recommend this as the default specification.

\begin{table}[ht]
\centering
\caption{Absolute (Abs) bias, 95\% credible interval coverage, and mean square error (MSE) in estimation of the population average causal effects using BART+SPL and BART applied to simulation 3.3B. RO-1 refers to simulations with the RO defined as $a=0.05*(range(X)),b=10$, RO-2 refers to simulations with the RO defined as $a=0.1*(range(X)),b=10$, and RO-3 refers to simulations with the RO defined as $a=0.15*(range(X)),b=3$.}
\begin{tabular}{rrrrrrr}
  \hline
$v,w$ & Method & $\pi$ & Abs Bias & Abs Bias (\%) & Coverage & MSE \\ 
  \hline
\multirow{4}{*}{$v=1.4,w=1.96$} & BART+SPL, RO-1 & 17 & 0.05 & 4.49 & 1.00 & 0.12 \\
& BART+SPL, RO-2 & 8 & 0.04 & 4.15 & 1.00 & 0.11 \\
& BART+SPL, RO-3 & 2 & 0.04 & 3.97 & 1.00 & 0.10 \\
& BART & & 0.05 & 4.69 & 0.57 & 0.12 \\ \hline
\multirow{4}{*}{$v=0.75,w=1.44$} &  BART+SPL, RO-1 & 24 & 0.07 & 6.27 & 1.00 & 0.14 \\
&  BART+SPL, RO-2 & 14 & 0.06 & 5.72 & 1.00 & 0.13 \\
&  BART+SPL, RO-3 & 6 & 0.06 & 5.83 & 1.00 & 0.12 \\
& BART & &  0.07 & 6.71 & 0.35 & 0.14 \\ \hline
\multirow{4}{*}{$v=0,w=1$}  & BART+SPL, RO-1 & 34 & 0.09 & 8.27 & 1.00 & 0.16 \\
&  BART+SPL, RO-2 & 21 & 0.06 & 5.72 & 1.00 & 0.12 \\
&  BART+SPL, RO-3 & 11 & 0.06 & 5.12 & 1.00 & 0.11 \\
& BART & & 0.06 & 5.67 & 0.74 & 0.12 \\ 
     \hline
\end{tabular}
\label{tab:chooseaB}
\end{table}

\section{The Effect of Natural Gas Compressor Stations on County-Level Thyroid Cancer and Leukemia Mortality}
\label{s:app}

We collected 2014 thyroid cancer and leukemia mortality rate estimates for each county in the US from the Global Health Data Exchange. The data and methods used to develop these estimates have been described previously \citep{mokdad2017trends}. We also obtained the locations of NG compressor stations from publicly available data complied by \cite{ornl2017}. While the data is not guaranteed to be complete, it is, to our knowledge, the most comprehensive documentation of compressor station locations in existence, with 1,359 compressor station locations verified using imagery. In order to test a causal hypothesis, we need to assume that exposure to compressor station-related emissions preceded 2014 (the year for which cancer mortality rates are observed) by at least the minimum latency period for thyroid cancer and leukemia. The CDC reports the minimum latency period for thyroid cancer as 2.5 years and the minimum latency for leukemia as 0.4 years \citep{wtc2015}. Although the dataset does not contain dates of origin for the compressor stations, it does contain peak operation dates. 84\% of the compressor stations in the dataset have peak operating dates in or before 2012; thus, it seems reasonable to assume that most of the compressor stations in the dataset operated at least 2.5 years prior to 2014. 

Our county-level exposure variable is an indicator of whether a compressor station is present in the county. We collected county-level demographic, socio-economic, and behavioral confounder data from the American Community Survey 2014 5-year estimates \citep{census2018} and the 2014 County Health Rankings and Roadmaps \citep{rwdf2018}. Data were accessed using Social Explorer. The confounders used are rate of primary care physicians, percent of less than 65 year olds uninsured, percent diabetic, percent current smokers, percent of people with limited access to healthy foods, percent obese, food environment index, population density, percent male, percent less than age 55, percent white, average household size, percent with bachelor's degree or higher, percent unemployed, median household income, Gini index of inequality, percent owner-occupied housing units, median rent as proportion of income, and average commute time to work. All the data used in this analysis are publicly available, and the data and R code to reproduce the analysis are posted on Github at \url{https://github.com/rachelnethery/overlap}.

We note that the sensitivity of this analysis to detect exposure effects will be low, because any true health effects of exposure to compression station emissions is likely more spatially concentrated than the county level. Although a higher spatial resolution analysis would be preferable, obtaining important behavioral confounder data at a finer spatial resolution is challenging. In an effort to improve the detectability of effects, we focus our analysis on roughly the mid-western region of the US (counties with centroid longitudes between -110 and -90), where few other sources of pollution exist compared to the coastal regions \citep{di2017air}. A focus on this region is also reasonable because NG production has a longer history in this region compared to other US production regions, likely leading to greater exposure.

\begin{figure}[h!]
\centering
\includegraphics[scale=.8]{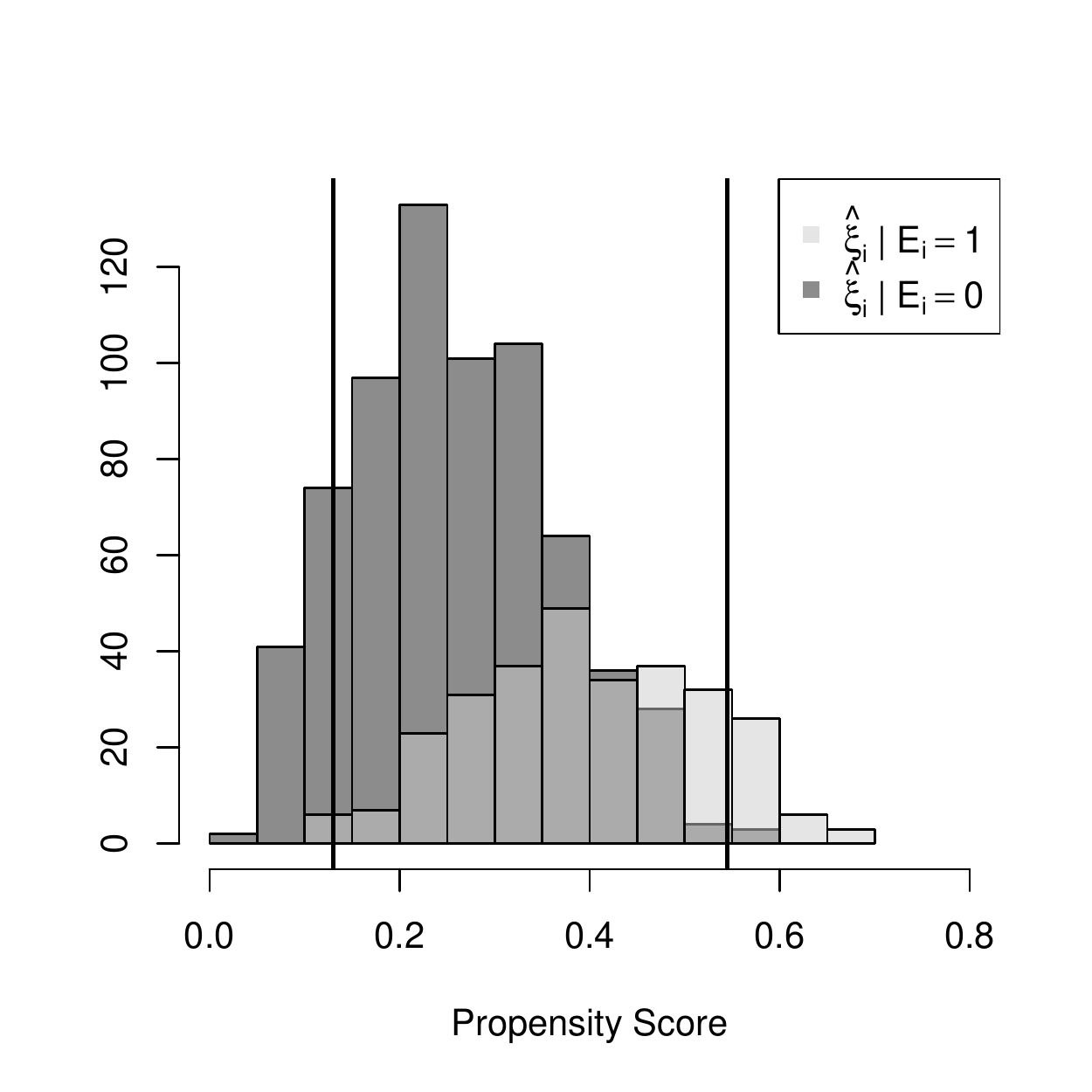}
\caption{Estimated propensity score histograms stratified by exposure status and overlaid. Bold vertical lines represent the start of non-overlap intervals in both tails of the distribution.}
\label{fig:psNG}
\end{figure}

We begin with a dataset of 1,309 counties, and, after discarding counties with any missing confounders, are left with N=978 counties. 291 of these counties are exposed (i.e., contain at least one compressor station) while 687 are unexposed. Table 7 in Section A.3 of the Appendix shows the differences in the exposed and unexposed populations. Notably, exposed counties have, on average, higher percent uninsured, lower population density, lower percent white, lower education, and higher percent unemployment. We estimate a propensity score by applying a BART probit with exposure status as the response and all the confounders as predictors. The histogram in Figure~\ref{fig:psNG} illustrates the non-overlap in the resulting propensity score, with the solid vertical lines denoting the start of the intervals of non-overlap that are detected in each tail of the propensity score using our overlap definition and $a=0.1*range(\hat{\xi})$ and $b=10$. With these specifications, 12\% of the sample falls into the RN. BART+SPL is needed to obtain population level causal effect estimates in this setting.

The following outcome variables are considered: (1) 2014 thyroid cancer mortality rate, (2) the change in thyroid cancer mortality rate from 1980 to 2014, (3) 2014 leukemia mortality rate, and (4) the change in leukemia mortality rate from 1980 to 2014. The 2014 rates are log-transformed prior to analysis. We analyze each outcome using both BART+SPL and trimmed BART. Counties in the trimmed sample are more urban and densely populated, on average, than the population represented by the full sample (trimmed sample average population density is 108.40 per mi$^2$ compared to 99.38 in the full sample).

Average causal effect estimates and 95\% credible intervals from each analysis can be found in Table~\ref{tab:aceNG}. The BART+SPL analysis is estimating population average causal effects, and the trimmed BART is estimating trimmed sample average causal effects. Only two of these analyses find statistically significant effects of NG compressor stations-- the trimmed BART analyses of the change in thyroid cancer mortality rates from 1980 to 2014 and of the change in leukemia mortality rates from 1980 to 2014. As discussed above, we must interpret these as significant effects only in the trimmed sample, which is on average more urban than the population of interest. The population-level estimates from BART+SPL have wider credible intervals for two reasons. First, the additional marginalization over the confounders required to obtain population-level estimates increases the variance. Second, to estimate at the population level, we must account for the additional uncertainty induced by the non-overlap, which BART+SPL does by inflating variances in the RN. With these wide credible intervals, evidence of an effect must be very compelling in order to achieve statistical significance. However, we note that the point estimate from each analysis is positive, indicating a harmful effect of compressor stations on health, and in each case the point estimates are quite similar in the trimmed and untrimmed analyses.

\begin{table}[ht]
\centering
\caption{Average causal effects of natural gas compressor station presence on 2014 county-level thyroid cancer and leukemia mortality rates and the change in thyroid cancer and leukemia mortality rates from 1980 to 2014.}
\begin{tabular}{rrrr}
  \hline
Outcome & Method & Effect & 95\% CI \\ 
  \hline
\multirow{2}{*}{2014 Thyroid Rates} & BART+SPL & 0.001 & -0.017, 0.020 \\ 
& BART & 0.003 & -0.007, 0.012 \\ \hline
\multirow{2}{*}{Change in Thyroid Rates 1980-2014} & BART+SPL & 0.992 & -0.308, 2.237 \\ 
& BART & 1.089 & 0.130, 2.038 \\ \hline
\multirow{2}{*}{2014 Leukemia Rates} & BART+SPL & 0.006 & -0.013, 0.025 \\ 
& BART & 0.005 & -0.004, 0.014\\ \hline
\multirow{2}{*}{Change in Leukemia Rates 1980-2014} & BART+SPL & 0.913 & -0.361, 2.206 \\
& BART & 0.988 & 0.014, 1.958 \\ 
   \hline
\end{tabular}
\label{tab:aceNG}
\end{table}

In Table 8 and Table 9 in Section A.3 of the Appendix, we provide the results of two sensitivity analyses using alternate specifications of $a$ and $b$, one resulting in a larger RN and one in a smaller RN. The BART+SPL results demonstrate little sensitivity to these choices, with point estimates changing little and inference remaining the same in each analysis. This robustness agrees with findings in simulated data in Section~\ref{ss:choosea}. The inference from the trimmed BART, however, is sensitive to the $a$ and $b$ choices, with the significant leukemia effect being attenuated in one sensitivity analysis and both significant effects being attenuated in the other. This sensitivity is not surprising, given that changes in the observations trimmed correspond to changes in the estimand. Because the trimmed BART is sensitive to these subjective choices, we should interpret the results with caution.

The significant and near-significant findings presented here suggest that the health effects of compressor station exposure is a topic that warrants further study with higher quality data. The data utilized here have numerous limitations that could be improved upon by future studies. In particular, an analysis at higher spatial resolution is needed in order to be able to detect geographically concentrated effects that may be washed out at the county level. Moreover, counties with compressor stations may also be more likely to be located in NG production regions; thus, an analysis at the county level may not be able to distinguish the effects of compressor station exposure from the effects of NG drilling and production-related exposures. Finally, an investigation of cancer diagnosis rates may be more informative about the possible dangers of NG compressor station exposure than our investigation of cancer mortality rates. However, cancer diagnosis rates are difficult to obtain across large geographic regions.

\section{Discussion}
\label{s:discuss}
In this paper, we have introduced a general definition of propensity score non-overlap and have proposed a Bayesian modeling approach to estimate population average causal effects and corresponding uncertainties in the presence of non-overlap. A novel feature of our proposed approach is its separation of the tasks of estimating causal effects in the region of overlap and the region of non-overlap and its delegation of these tasks to two distinct models. A BART model is selected to perform estimation of individual causal effects in the region of overlap where there is strong data support, thanks to its non-parametric nature, its ability to capture heterogeneous effects, and its strong predictive capacity. In the region of non-overlap, where reliance on model specification is required to estimate causal effects, individual causal effects are estimated by extrapolating trends from the region of overlap via a parametric spline model. This approach, which we call BART+SPL, can be applied to data with either continuous or binary outcomes and can be implemented in a fully Bayesian manner, so that uncertainties from both stages are captured. 

We demonstrated via simulations that BART+SPL outperforms both stand-alone BART and stand-alone spline causal inference approaches in estimation of population average causal effects under a wide range of conditions involving propensity score non-overlap. However, due to BART's limitations in high dimensional settings, BART+SPL may give more biased results than existing methods when many irrelevant predictors are present.

While we have focused primarily on the use of our overlap definition with BART+SPL, it can also be used to define the RO for trimming, and it may provide a more transparent and reproducible approach than trimming by eye. The rich causal inference literature on caliper selection for matching procedures may provide insight on how to specify $a$ and $b$ for the use of this overlap definition with trimming \citep{stuart_matching_2010,lunt_selecting_2014}. However, as demonstrated by the trimmed BART results in Section~\ref{s:app}, our overlap definition is unlikely to produce an interpretable trimmed estimand. Therefore, other strategies that prioritize the interpretability of the resulting estimand may be preferable, such as overlap weighting \citep{li_balancing_2016}.

We again note that BART+SPL is intended to handle finite sample non-overlap only, because the appropriateness of estimating population average causal effects may be called into question when the population exhibits non-overlap. When non-overlap is a finite sample feature, it will disappear asymptotically. It is for this reason that we have avoided discussions of double robustness. Because double robustness is an asymptotic property, and BART+SPL is a finite sample method, we cannot make claims about double robustness in the classic sense. However, we have demonstrated in simulations that, relative to the competing methods considered, BART+SPL is most robust to both outcome and propensity score model misspecification in the presence of non-overlap.

A key contribution of our work is the introduction of a tuning parameter used to inflate the variance of causal effect estimates in regions of poor data support, so that this variance adequately reflects the high estimation uncertainty in such regions. We have recommended a choice of tuning parameter that linearly increases the variance as the distance into the RN increases, and the scale of these increases is tailored to correspond reasonably to the scale of the estimated causal effects in regions of good data support. Simulation results demonstrated that the resulting credible intervals are much more reliable than those produced by competing methods in the presence of non-overlap. However, in ``simple'' scenarios (i.e., scenarios where trends in the causal effects in the RN are easily predictable based on trends in the RO), this tuning parameter can produce conservative uncertainties.

Although to our knowledge, the use of Gaussian Process Regression (GPR) for causal inference has not previously been discussed in the literature, there are clear connections between the features of our approach and the features of GPR, thus a comparison of the two methods is warranted. GPR is a traditionally Bayesian non-parametric regression technique that employs kernel functions to interpolate and predict the outcome at unobserved points. It naturally accommodates non-linearities and identifies regions of poor data support and inflates uncertainties in those regions.

While these features might make it an attractive and parsimonious approach to causal inference in the presence of non-overlap, BART+SPL provides the following three advantages over GPR that may render it more appealing, particularly to applied scientists: (1) less sensitivity to tuning choices, (2) more intuitive tuning parameters, and (3) greater computational scalability. BART is highly regarded for its strong and consistent performance under default tuning specifications, and, in Section~\ref{ss:choosea}, BART+SPL also demonstrated little sensitivity to tuning choices. The results of GPR are known to be highly sensitive to the choice of kernel and tuning parameters. Similarly, GPR's tuning parameters are often difficult to understand, as they are embedded in kernel functions within covariance matrices; therefore, tuning typically requires guess-and-check work. BART+SPL involves tuning parameters with straightforward interpretations relating to the definition of adequate data support. Finally, GPR requires manipulation of a $N \times N$ covariance matrix in each MCMC iteration, making it time-consuming or infeasible with large datasets, while BART+SPL is much more scalable.

Our work introduces an exciting direction for methodological developments in the context of causal inference with propensity score non-overlap. For instance, other machine learning methods \citep{cristianini2000introduction,breiman2001random,schmidhuber2015deep} may also have properties that make them well-suited to handle non-overlap, and the replacement of BART with one of these other methods could be explored. Moreover, the limitations of BART in high dimensions provide an opportunity for improvement on our method. Improvements might be made through the use of other machine learning methods, through the integration of sparsity-inducing priors with BART as developed by \cite{linero_bayesian_2016}, or through the addition of pre-processing procedures to first identify a minimal confounder set. In the spirit of Bayesian Adjustment for Confounding \citep{wang2012bayesian}, propensity score model and outcome model variable selection could be accomplished simultaneously. For instance, a propensity score model and a BART+SPL model could be fit jointly, both utilizing sparsity inducing priors, where the propensity score model priors are informed by the strength of the relationship between the outcome and each covariate, and the BART+SPL model priors are informed by the strength of the relationship between the exposure and each covariate. Finally, theoretical results remain to be developed for methods that perform causal effect estimation in the RO and RN using distinct models. These could be avenues for future work.

\section{Acknowledgements}

Support for this work was provided by NIH grants 5T32ES007142-35, R01ES028033, P01CA134294, R01GM111339, R35CA197449, R01ES026217, P50MD010428, and R01MD012769. The authors also received support from EPA grants 83615601 and 83587201-0, and Health Effects Institute grant 4953-RFA14-3/16-4.

\bibliographystyle{Chicago}
\bibliography{overlap_bib}

\begin{thebibliography}{}

\bibitem[\protect\citeauthoryear{Austin}{Austin}{2011}]{austin_introduction_2011}
Austin, P.~C. (2011).
\newblock An introduction to propensity score methods for reducing the effects
  of confounding in observational studies.
\newblock {\em Multivariate Behavioral Research\/}~{\em 46\/}(3), 399--424.

\bibitem[\protect\citeauthoryear{Bonato, Baladandayuthapani, Broom, Sulman,
  Aldape, and Do}{Bonato et~al.}{2011}]{bonato2011bayesian}
Bonato, V., V.~Baladandayuthapani, B.~M. Broom, E.~P. Sulman, K.~D. Aldape, and
  K.-A. Do (2011).
\newblock Bayesian ensemble methods for survival prediction in gene expression
  data.
\newblock {\em Bioinformatics\/}~{\em 27\/}(3), 359--367.

\bibitem[\protect\citeauthoryear{Breiman}{Breiman}{2001}]{breiman2001random}
Breiman, L. (2001).
\newblock Random forests.
\newblock {\em Machine Learning\/}~{\em 45\/}(1), 5--32.

\bibitem[\protect\citeauthoryear{Chipman, George, and McCulloch}{Chipman
  et~al.}{1998}]{chipman_bayesian_1998}
Chipman, H.~A., E.~I. George, and R.~E. McCulloch (1998).
\newblock Bayesian {CART} {Model} {Search}.
\newblock {\em Journal of the American Statistical Association\/}~{\em
  93\/}(443), 935--948.

\bibitem[\protect\citeauthoryear{Chipman, George, and McCulloch}{Chipman
  et~al.}{2010}]{chipman_bart_2010}
Chipman, H.~A., E.~I. George, and R.~E. McCulloch (2010).
\newblock {BART}: {Bayesian} additive regression trees.
\newblock {\em The Annals of Applied Statistics\/}~{\em 4\/}(1), 266--298.

\bibitem[\protect\citeauthoryear{Cochran and Rubin}{Cochran and
  Rubin}{1973}]{cochran1973controlling}
Cochran, W.~G. and D.~B. Rubin (1973).
\newblock Controlling bias in observational studies: A review.
\newblock {\em Sankhy{\=a}: The Indian Journal of Statistics, Series A\/}~{\em
  35\/}(4), 417--446.

\bibitem[\protect\citeauthoryear{Cole and Hern\'an}{Cole and
  Hern\'an}{2008}]{cole_constructing_2008}
Cole, S.~R. and M.~A. Hern\'an (2008).
\newblock Constructing inverse probability weights for marginal structural
  models.
\newblock {\em American Journal of Epidemiology\/}~{\em 168\/}(6), 656--664.

\bibitem[\protect\citeauthoryear{Cristianini and Shawe-Taylor}{Cristianini and
  Shawe-Taylor}{2000}]{cristianini2000introduction}
Cristianini, N. and J.~Shawe-Taylor (2000).
\newblock {\em An Introduction to Support Vector Machines and Other
  Kernel-Based Learning Methods}.
\newblock Cambridge University Press.

\bibitem[\protect\citeauthoryear{Crump, Hotz, Imbens, and Mitnik}{Crump
  et~al.}{2009}]{crump_dealing_2009}
Crump, R.~K., V.~J. Hotz, G.~W. Imbens, and O.~A. Mitnik (2009).
\newblock Dealing with limited overlap in estimation of average treatment
  effects.
\newblock {\em Biometrika\/}~{\em 96\/}(1), 187--199.

\bibitem[\protect\citeauthoryear{D'Amour, Deng, Feller, Lei, and
  Sekhon}{D'Amour et~al.}{2017}]{damour2017overlap}
D'Amour, A., P.~Deng, A.~Feller, L.~Lei, and J.~Sekhon (2017).
\newblock Overlap in observational studies with high-dimensional covariates.
\newblock {\em arXiv preprint arXiv:1711.02582\/}.

\bibitem[\protect\citeauthoryear{Dehejia and Wahba}{Dehejia and
  Wahba}{1999}]{dehejia1999causal}
Dehejia, R.~H. and S.~Wahba (1999).
\newblock Causal effects in nonexperimental studies: Reevaluating the
  evaluation of training programs.
\newblock {\em Journal of the American Statistical Association\/}~{\em
  94\/}(448), 1053--1062.

\bibitem[\protect\citeauthoryear{Di, Wang, Zanobetti, Wang, Koutrakis, Choirat,
  Dominici, and Schwartz}{Di et~al.}{2017}]{di2017air}
Di, Q., Y.~Wang, A.~Zanobetti, Y.~Wang, P.~Koutrakis, C.~Choirat, F.~Dominici,
  and J.~D. Schwartz (2017).
\newblock Air pollution and mortality in the medicare population.
\newblock {\em New England Journal of Medicine\/}~{\em 376\/}(26), 2513--2522.

\bibitem[\protect\citeauthoryear{Ferguson, McElrath, and Meeker}{Ferguson
  et~al.}{2014}]{ferguson2014environmental}
Ferguson, K.~K., T.~F. McElrath, and J.~D. Meeker (2014).
\newblock Environmental phthalate exposure and preterm birth.
\newblock {\em JAMA Pediatrics\/}~{\em 168\/}(1), 61--68.

\bibitem[\protect\citeauthoryear{Finkel}{Finkel}{2016}]{finkel2016shale}
Finkel, M. (2016).
\newblock Shale gas development and cancer incidence in southwest
  {P}ennsylvania.
\newblock {\em Public Health\/}~{\em 141}, 198--206.

\bibitem[\protect\citeauthoryear{Golding and Watson}{Golding and
  Watson}{1999}]{golding1999possible}
Golding, B. and W.~Watson (1999).
\newblock Possible mechanisms of carcinogenesis after exposure to benzene.
\newblock {\em IARC Scientific Publications\/}~(150), 75--88.

\bibitem[\protect\citeauthoryear{Gutman and Rubin}{Gutman and
  Rubin}{2015}]{gutman2015estimation}
Gutman, R. and D.~B. Rubin (2015).
\newblock Estimation of causal effects of binary treatments in unconfounded
  studies.
\newblock {\em Statistics in Medicine\/}~{\em 34\/}(26), 3381--3398.

\bibitem[\protect\citeauthoryear{Hastie and Tibshirani}{Hastie and
  Tibshirani}{2000}]{hastie2000bayesian}
Hastie, T. and R.~Tibshirani (2000).
\newblock Bayesian backfitting.
\newblock {\em Statistical Science\/}~{\em 15\/}(3), 196--223.

\bibitem[\protect\citeauthoryear{He, Li, Viant, and Yao}{He
  et~al.}{2009}]{he2009profiling}
He, S., X.~Li, M.~R. Viant, and X.~Yao (2009).
\newblock Profiling {MS} proteomics data using smoothed non-linear energy
  operator and {B}ayesian additive regression trees.
\newblock {\em Proteomics\/}~{\em 9\/}(17), 4176--4191.

\bibitem[\protect\citeauthoryear{Hill and Su}{Hill and
  Su}{2013}]{hill2013assessing}
Hill, J. and Y.-S. Su (2013).
\newblock Assessing lack of common support in causal inference using bayesian
  nonparametrics: Implications for evaluating the effect of breastfeeding on
  children's cognitive outcomes.
\newblock {\em The Annals of Applied Statistics\/}, 1386--1420.

\bibitem[\protect\citeauthoryear{Hill}{Hill}{2011}]{hill_bayesian_2011}
Hill, J.~L. (2011).
\newblock Bayesian nonparametric modeling for causal inference.
\newblock {\em Journal of Computational and Graphical Statistics\/}~{\em
  20\/}(1), 217--240.

\bibitem[\protect\citeauthoryear{Ho, Imai, King, and Stuart}{Ho
  et~al.}{2007}]{ho2007matching}
Ho, D.~E., K.~Imai, G.~King, and E.~A. Stuart (2007).
\newblock Matching as nonparametric preprocessing for reducing model dependence
  in parametric causal inference.
\newblock {\em Political Analysis\/}~{\em 15\/}(3), 199--236.

\bibitem[\protect\citeauthoryear{Huang, Marco, Pinello, and Yuan}{Huang
  et~al.}{2015}]{huang2015predicting}
Huang, J., E.~Marco, L.~Pinello, and G.-C. Yuan (2015).
\newblock Predicting chromatin organization using histone marks.
\newblock {\em Genome Biology\/}~{\em 16\/}(1), 162.

\bibitem[\protect\citeauthoryear{Kassotis, Tillitt, Lin, McElroy, and
  Nagel}{Kassotis et~al.}{2016}]{kassotis2016endocrine}
Kassotis, C.~D., D.~E. Tillitt, C.-H. Lin, J.~A. McElroy, and S.~C. Nagel
  (2016).
\newblock Endocrine-disrupting chemicals and oil and natural gas operations:
  {P}otential environmental contamination and recommendations to assess complex
  environmental mixtures.
\newblock {\em Environmental Health Perspectives\/}~{\em 124\/}(3), 256.

\bibitem[\protect\citeauthoryear{Kindo, Wang, and Pe{\~n}a}{Kindo
  et~al.}{2016}]{kindo2016multinomial}
Kindo, B.~P., H.~Wang, and E.~A. Pe{\~n}a (2016).
\newblock Multinomial probit {B}ayesian additive regression trees.
\newblock {\em Stat\/}~{\em 5\/}(1), 119--131.

\bibitem[\protect\citeauthoryear{King and Zeng}{King and
  Zeng}{2005}]{king2005dangers}
King, G. and L.~Zeng (2005).
\newblock The dangers of extreme counterfactuals.
\newblock {\em Political Analysis\/}~{\em 14\/}(2), 131--159.

\bibitem[\protect\citeauthoryear{Kloczko}{Kloczko}{2015}]{EHP2015-2}
Kloczko, N. (2015).
\newblock Summary on compressor stations and health impacts.
\newblock
  \url{http://www.environmentalhealthproject.org/files/A%20Brief%20Review%20of%20Compressor%20Stations%2011.2015.pdf}.
\newblock Accessed: 2018-03-27.

\bibitem[\protect\citeauthoryear{LaLonde}{LaLonde}{1986}]{lalonde1986evaluating}
LaLonde, R.~J. (1986).
\newblock Evaluating the econometric evaluations of training programs with
  experimental data.
\newblock {\em The American Economic Review\/}~{\em 76\/}(4), 604--620.

\bibitem[\protect\citeauthoryear{Li, Morgan, and Zaslavsky}{Li
  et~al.}{018a}]{li_balancing_2016}
Li, F., K.~L. Morgan, and A.~M. Zaslavsky (2018a).
\newblock Balancing covariates via propensity score weighting.
\newblock {\em Journal of the American Statistical Association\/}~{\em
  113\/}(521), 390--400.

\bibitem[\protect\citeauthoryear{Li, Thomas, and Li}{Li
  et~al.}{018b}]{li2018addressing}
Li, F., L.~E. Thomas, and F.~Li (2018b).
\newblock Addressing extreme propensity scores via the overlap weights.
\newblock {\em American Journal of Epidemiology\/}, kwy201.

\bibitem[\protect\citeauthoryear{Linero}{Linero}{2016}]{linero_bayesian_2016}
Linero, A.~R. (2016).
\newblock Bayesian regression trees for high dimensional prediction and
  variable selection.
\newblock {\em Journal of the American Statistical Association\/}.

\bibitem[\protect\citeauthoryear{Liu, Shao, and Yuan}{Liu
  et~al.}{2010}]{liu2010prediction}
Liu, Y., Z.~Shao, and G.-C. Yuan (2010).
\newblock Prediction of polycomb target genes in mouse embryonic stem cells.
\newblock {\em Genomics\/}~{\em 96\/}(1), 17--26.

\bibitem[\protect\citeauthoryear{Liu, Traskin, Lorch, George, and Small}{Liu
  et~al.}{2015}]{liu2015ensemble}
Liu, Y., M.~Traskin, S.~A. Lorch, E.~I. George, and D.~Small (2015).
\newblock Ensemble of trees approaches to risk adjustment for evaluating a
  hospital’s performance.
\newblock {\em Health Care Management Science\/}~{\em 18\/}(1), 58--66.

\bibitem[\protect\citeauthoryear{Lunt}{Lunt}{2014}]{lunt_selecting_2014}
Lunt, M. (2014).
\newblock Selecting an appropriate caliper can be essential for achieving good
  balance with propensity score matching.
\newblock {\em American Journal of Epidemiology\/}~{\em 179\/}(2), 226--235.

\bibitem[\protect\citeauthoryear{Maltoni, Ciliberti, Cotti, Conti, and
  Belpoggi}{Maltoni et~al.}{1989}]{maltoni1989benzene}
Maltoni, C., A.~Ciliberti, G.~Cotti, B.~Conti, and F.~Belpoggi (1989).
\newblock Benzene, an experimental multipotential carcinogen: {R}esults of the
  long-term bioassays performed at the {B}ologna {I}nstitute of {O}ncology.
\newblock {\em Environmental Health Perspectives\/}~{\em 82}, 109.

\bibitem[\protect\citeauthoryear{McKenzie, Allshouse, Byers, Bedrick, Serdar,
  and Adgate}{McKenzie et~al.}{2017}]{mckenzie2017childhood}
McKenzie, L.~M., W.~B. Allshouse, T.~E. Byers, E.~J. Bedrick, B.~Serdar, and
  J.~L. Adgate (2017).
\newblock Childhood hematologic cancer and residential proximity to oil and gas
  development.
\newblock {\em PLOS One\/}~{\em 12\/}(2), e0170423.

\bibitem[\protect\citeauthoryear{Messersmith, Brockett, and
  Loveland}{Messersmith et~al.}{2015}]{messersmith2015}
Messersmith, D., D.~Brockett, and C.~Loveland (2015).
\newblock Understanding natural gas compressor stations.
\newblock {\em Penn State Extension\/}.

\bibitem[\protect\citeauthoryear{Mokdad, Dwyer-Lindgren, Fitzmaurice, Stubbs,
  Bertozzi-Villa, Morozoff, Charara, Allen, Naghavi, and Murray}{Mokdad
  et~al.}{2017}]{mokdad2017trends}
Mokdad, A.~H., L.~Dwyer-Lindgren, C.~Fitzmaurice, R.~W. Stubbs,
  A.~Bertozzi-Villa, C.~Morozoff, R.~Charara, C.~Allen, M.~Naghavi, and C.~J.
  Murray (2017).
\newblock Trends and patterns of disparities in cancer mortality among {US}
  counties, 1980-2014.
\newblock {\em Journal of the American Medical Association\/}~{\em 317\/}(4),
  388--406.

\bibitem[\protect\citeauthoryear{{Oak Ridge National Laboratory}}{{Oak Ridge
  National Laboratory}}{2017}]{ornl2017}
{Oak Ridge National Laboratory} (2017).
\newblock Natural gas compressor stations.
\newblock
  \url{https://hifld-dhs-gii.opendata.arcgis.com/datasets/fd7d62905d194eba87d2ee18d1a244b3_0}.
\newblock Accessed: 2018-03-29.

\bibitem[\protect\citeauthoryear{Pellegriti, Frasca, Regalbuto, Squatrito, and
  Vigneri}{Pellegriti et~al.}{2013}]{pellegriti2013worldwide}
Pellegriti, G., F.~Frasca, C.~Regalbuto, S.~Squatrito, and R.~Vigneri (2013).
\newblock Worldwide increasing incidence of thyroid cancer: {U}pdate on
  epidemiology and risk factors.
\newblock {\em Journal of Cancer Epidemiology\/}~{\em 2013}.

\bibitem[\protect\citeauthoryear{{Pennsylvania Department of Environmental
  Protection}}{{Pennsylvania Department of Environmental
  Protection}}{2010}]{padep2010}
{Pennsylvania Department of Environmental Protection} (2010).
\newblock Southwestern {P}ennsylvania {M}arcellus {S}hale short-term ambient
  air sampling report.
\newblock
  \url{http://www.dep.state.pa.us/dep/deputate/airwaste/aq/aqm/docs/Marcellus_SW_11-01-10.pdf}.
\newblock Accessed: 2018-03-29.

\bibitem[\protect\citeauthoryear{Petersen, Porter, Gruber, Wang, and van~der
  Laan}{Petersen et~al.}{2012}]{petersen_diagnosing_2012}
Petersen, M.~L., K.~E. Porter, S.~Gruber, Y.~Wang, and M.~J. van~der Laan
  (2012).
\newblock Diagnosing and responding to violations in the positivity assumption.
\newblock {\em Statistical Methods in Medical Research\/}~{\em 21\/}(1),
  31--54.

\bibitem[\protect\citeauthoryear{{R Core Team}}{{R Core Team}}{2016}]{R2016}
{R Core Team} (2016).
\newblock {\em R: A Language and Environment for Statistical Computing}.
\newblock Vienna, Austria: R Foundation for Statistical Computing.

\bibitem[\protect\citeauthoryear{Rasmussen, Ogburn, McCormack, Casey,
  Bandeen-Roche, Mercer, and Schwartz}{Rasmussen
  et~al.}{2016}]{rasmussen2016association}
Rasmussen, S.~G., E.~L. Ogburn, M.~McCormack, J.~A. Casey, K.~Bandeen-Roche,
  D.~G. Mercer, and B.~S. Schwartz (2016).
\newblock Association between unconventional natural gas development in the
  {M}arcellus {S}hale and asthma exacerbations.
\newblock {\em JAMA Internal Medicine\/}~{\em 176\/}(9), 1334--1343.

\bibitem[\protect\citeauthoryear{{Robert Wood Johnson Foundation}}{{Robert Wood
  Johnson Foundation}}{2014}]{rwdf2018}
{Robert Wood Johnson Foundation} (2014).
\newblock 2014 county health rankings and roadmaps.
\newblock Prepared by Social Explorer.
\newblock Accessed: 2018-03-27.

\bibitem[\protect\citeauthoryear{Rosenbaum and Rubin}{Rosenbaum and
  Rubin}{1983}]{rosenbaum_central_1983}
Rosenbaum, P.~R. and D.~B. Rubin (1983).
\newblock The central role of the propensity score in observational studies for
  causal effects.
\newblock {\em Biometrika\/}~{\em 70\/}(1), 41--55.

\bibitem[\protect\citeauthoryear{Rubin}{Rubin}{1974}]{rubin_estimating_1974}
Rubin, D.~B. (1974).
\newblock Estimating causal effects of treatments in randomized and
  nonrandomized studies.
\newblock {\em Journal of Educational Psychology\/}~{\em 66\/}(5), 688--701.

\bibitem[\protect\citeauthoryear{Rubin}{Rubin}{1980}]{rubinrandomization1980}
Rubin, D.~B. (1980).
\newblock Randomization analysis of experimental data: {The} {Fisher}
  randomization test comment.
\newblock {\em Journal of the American Statistical Association\/}~{\em
  75\/}(371), 591--593.

\bibitem[\protect\citeauthoryear{Schmidhuber}{Schmidhuber}{2015}]{schmidhuber2015deep}
Schmidhuber, J. (2015).
\newblock Deep learning in neural networks: An overview.
\newblock {\em Neural Networks\/}~{\em 61}, 85--117.

\bibitem[\protect\citeauthoryear{{Southwest Pennsylvania Environmental Health
  Project}}{{Southwest Pennsylvania Environmental Health
  Project}}{2015}]{EHP2015}
{Southwest Pennsylvania Environmental Health Project} (2015).
\newblock Summary on compressor stations and health impacts.
\newblock
  \url{http://www.environmentalhealthproject.org/files/Summary%20Compressor-station-emissions-and-health-impacts-02.24.2015.pdf}.
\newblock Accessed: 2018-03-27.

\bibitem[\protect\citeauthoryear{Sparapani, Logan, McCulloch, and
  Laud}{Sparapani et~al.}{2016}]{sparapani2016nonparametric}
Sparapani, R.~A., B.~R. Logan, R.~E. McCulloch, and P.~W. Laud (2016).
\newblock Nonparametric survival analysis using {B}ayesian additive regression
  trees {(BART)}.
\newblock {\em Statistics in Medicine\/}~{\em 35\/}(16), 2741--2753.

\bibitem[\protect\citeauthoryear{Strobl, Malley, and Tutz}{Strobl
  et~al.}{2009}]{strobl2009introduction}
Strobl, C., J.~Malley, and G.~Tutz (2009).
\newblock An introduction to recursive partitioning: rationale, application,
  and characteristics of classification and regression trees, bagging, and
  random forests.
\newblock {\em Psychological Methods\/}~{\em 14\/}(4), 323.

\bibitem[\protect\citeauthoryear{Stuart}{Stuart}{2010}]{stuart_matching_2010}
Stuart, E.~A. (2010).
\newblock Matching methods for causal inference: {A} review and a look forward.
\newblock {\em Statistical Science: A Review Journal of the Institute of
  Mathematical Statistics\/}~{\em 25\/}(1), 1--21.

\bibitem[\protect\citeauthoryear{{US Census Bureau}}{{US Census
  Bureau}}{2014}]{census2018}
{US Census Bureau} (2014).
\newblock American community survey 2014 (5 year estimates).
\newblock Prepared by Social Explorer.
\newblock Accessed: 2018-03-27.

\bibitem[\protect\citeauthoryear{{US EPA}}{{US EPA}}{2018}]{epaEDSP}
{US EPA} (2018).
\newblock Endocrine disruptor screening program {(EDSP)} estrogen receptor
  bioactivity.
\newblock
  \url{https://www.epa.gov/endocrine-disruption/endocrine-disruptor-screening-program-edsp-estrogen-receptor-bioactivity#main-content}.
\newblock Accessed: 2018-03-27.

\bibitem[\protect\citeauthoryear{Wang, Dominici, Parmigiani, and Zigler}{Wang
  et~al.}{2015}]{wang2015accounting}
Wang, C., F.~Dominici, G.~Parmigiani, and C.~M. Zigler (2015).
\newblock Accounting for uncertainty in confounder and effect modifier
  selection when estimating average causal effects in generalized linear
  models.
\newblock {\em Biometrics\/}~{\em 71\/}(3), 654--665.

\bibitem[\protect\citeauthoryear{Wang, Parmigiani, and Dominici}{Wang
  et~al.}{2012}]{wang2012bayesian}
Wang, C., G.~Parmigiani, and F.~Dominici (2012).
\newblock Bayesian effect estimation accounting for adjustment uncertainty.
\newblock {\em Biometrics\/}~{\em 68\/}(3), 661--671.

\bibitem[\protect\citeauthoryear{Westreich and Cole}{Westreich and
  Cole}{2010}]{westreich_invited_2010}
Westreich, D. and S.~R. Cole (2010).
\newblock Invited commentary: Positivity in practice.
\newblock {\em American Journal of Epidemiology\/}~{\em 171\/}(6), 674--677.

\bibitem[\protect\citeauthoryear{Westreich, Lessler, and Funk}{Westreich
  et~al.}{2010}]{westreich2010propensity}
Westreich, D., J.~Lessler, and M.~J. Funk (2010).
\newblock Propensity score estimation: neural networks, support vector
  machines, decision trees {(CART)}, and meta-classifiers as alternatives to
  logistic regression.
\newblock {\em Journal of Clinical Epidemiology\/}~{\em 63\/}(8), 826--833.

\bibitem[\protect\citeauthoryear{{Wolf Eagle Environmental}}{{Wolf Eagle
  Environmental}}{2009}]{wolfeagle2009}
{Wolf Eagle Environmental} (2009).
\newblock Town of {Dish, Texas} ambient air monitoring analysis final report.
\newblock \url{https://townofdish.com/objects/DISH_-_final_report_revised.pdf}.
\newblock Accessed: 2018-05-08.

\bibitem[\protect\citeauthoryear{{World Trade Center Health Program}}{{World
  Trade Center Health Program}}{2015}]{wtc2015}
{World Trade Center Health Program} (2015).
\newblock Minimum latency \& types or categories of cancer.
\newblock
  \url{https://www.cdc.gov/wtc/pdfs/WTCHP-Minimum-Cancer-Latency-PP-01062015.pdf}.
\newblock Accessed: 2018-03-29.

\bibitem[\protect\citeauthoryear{Yang and Ding}{Yang and
  Ding}{2018}]{yang2018asymptotic}
Yang, S. and P.~Ding (2018).
\newblock Asymptotic inference of causal effects with observational studies
  trimmed by the estimated propensity scores.
\newblock {\em Biometrika\/}.

\bibitem[\protect\citeauthoryear{Zhou and Liu}{Zhou and
  Liu}{2008}]{zhou2008extracting}
Zhou, Q. and J.~S. Liu (2008).
\newblock Extracting sequence features to predict protein--dna interactions: a
  comparative study.
\newblock {\em Nucleic Acids Research\/}~{\em 36\/}(12), 4137--4148.

\bibitem[\protect\citeauthoryear{Zigler, Kim, Choirat, Hansen, Wang, Hund,
  Samet, King, and Dominici}{Zigler et~al.}{2016}]{zigler2016hei}
Zigler, C., C.~Kim, C.~Choirat, J.~Hansen, Y.~Wang, L.~Hund, J.~Samet, G.~King,
  and F.~Dominici (2016).
\newblock Causal inference methods for estimating long-term health effects of
  air quality regulations. {R}esearch report 187.
\newblock {\em Boston, MA: Health Effects Institute\/}.

\bibitem[\protect\citeauthoryear{Zigler and Cefalu}{Zigler and
  Cefalu}{2017}]{zigler2017posterior}
Zigler, C.~M. and M.~Cefalu (2017).
\newblock Posterior predictive treatment assignment for estimating causal
  effects with limited overlap.
\newblock {\em arXiv preprint arXiv:1710.08749\/}.

\end{thebibliography}

\clearpage

\appendix

\section{Appendix}

\subsection{BART+SPL MCMC Sampling Scheme for Continuous Outcomes}
\label{as:mcmc}
BART+SPL can be implemented via an MCMC sampling scheme that is easily executed in R software \citep{R2016} with the help of the existing dbarts package \citep{chipman_bart_2010} for drawing samples from a BART. The full conditional distribution of the parameters in a given tree in the BART,\\$p(\mathcal{T}_j, \mathcal{M}_j | \mathbf{Y}_O^{obs},\sigma^2_B, \mathcal{T}_1,...,\mathcal{T}_{j-1},\mathcal{T}_{j+1},...,\mathcal{T}_J,\mathcal{M}_1,...,\mathcal{M}_{j-1},\mathcal{M}_{j+1},...,\mathcal{M}_J)$, depends only on the observed data and the other tree parameters only through the partial residuals $\boldsymbol{V}_{Oj}=\mathbf{Y}_O^{obs}-\sum_{l \neq j} g(\mathbf{X}_{O};\mathcal{T}_l,\mathcal{M}_l)$. Invoking this simplification, BART can be fit using a technique called Bayesian backfitting or backfitting MCMC \citep{hastie2000bayesian}, where the MCMC samples from a given tree are drawn while treating the parameters from all other trees as constant. We outline the steps for obtaining MCMC samples from BART+SPL below, assuming the default BART prior distributions recommended in \citet{chipman_bart_2010} and non-informative Normal-Inverse Gamma priors for the spline model. This sampling scheme assumes that only one smoothing stage model is fit, i.e., the RN contains observations from only one exposure group; however, if the RN contains observations from both exposure groups and two splines are needed, steps (5) and (6) can easily be repeated for a second spline model. We assume that each of the BART and spline parameters has been initialized, so that we begin with $\left\lbrace \mathcal{T}_j^{(0)},\mathcal{M}_j^{(0)},{\sigma_B^2}^{(0)},\boldsymbol{\beta}^{(0)},{\sigma^2_S}^{(0)} \right\rbrace$.

{\noindent For $m=1,...,M$,}
\begin{enumerate}

\item For tree $j$, ($j=1,...,J$)
\begin{enumerate}

\item Draw $\mathcal{T}_j^{(m)}$ from $p(\mathcal{T}_j|\boldsymbol{V}_{Oj}^{(m-1)},{\sigma^2_B}^{(m-1)})$ using the Metropolis Hastings algorithm described by \citet{chipman_bayesian_1998}

\item Draw $\mathcal{M}_j^{(m)}$ from $p(\mathcal{M}_j|\boldsymbol{V}_{Oj}^{(m-1)},{\sigma^2_B}^{(m-1)},\mathcal{T}_j^{(m)})$ through a random sample from the Normal distribution
\end{enumerate}

\item Draw ${\sigma_B^2}^{(m)}$ from $p({\sigma_B^2}|\mathbf{Y}_O^{obs}, \mathcal{T}_1^{(m)},...,\mathcal{T}_J^{(m)},\mathcal{M}_1^{(m)},...,\mathcal{M}_J^{(m)})$ through a random sample from an Inverse-Gamma distribution

\item Draw ${\mathbf{Y}_O^{mis}}^{(m)}$ from $p(\mathbf{Y}_O^{mis}|\mathbf{Y}_O^{obs}, \mathcal{T}_1^{(m)},...,\mathcal{T}_J^{(m)},\mathcal{M}_1^{(m)},...,\mathcal{M}_J^{(m)},{\sigma^2_B}^{(m)})$ through a random sample from a Normal distribution

\item Form $\mathbf{\Delta}_O^{(m)}$ as the appropriate linear combination of $\mathbf{Y}_O^{obs}$ and ${\mathbf{Y}_O^{mis}}^{(m)}$

\item Draw $\boldsymbol{\beta}^{(m)}$ from $p(\boldsymbol{\beta}|\mathbf{\Delta}_O^{(m)},{\sigma^2_S}{(m-1)})$ through a random sample from the Normal distribution

\item Draw ${\sigma_S^2}^{(m)}$ from $p(\sigma_S^2|\mathbf{\Delta}_O^{(m)},\boldsymbol{\beta}^{(m)})$ through a random sample from an Inverse-Gamma distribution

\item Draw $\mathbf{\Delta}_{O^\perp}^{(m)}$ from $p(\mathbf{\Delta}_{O^\perp}|\mathbf{\Delta}_O^{(m)},\boldsymbol{\beta}^{(m)},{\sigma_S^2}^{(m)},\tau_{q})$ through a random sample from a Normal distribution

\item Draw $\Delta^{(m)}_P$ by executing $B$ iterations of the Bayesian bootstrap on $\left\lbrace \tilde{\mathbf{\Delta}}^{(m)}_{O}, \tilde{\mathbf{\Delta}}^{(m)}_{O^\perp} \right\rbrace$ and randomly selecting one of the $B$ bootstrap sample averages
\end{enumerate}

\clearpage

\subsection{Simulations with Binary Outcomes}

In simulations involving a binary outcome, we again consider scenarios using the true propensity score (2A) and simulations using a misspecified propensity score estimate (2B). In 2A, we let $N$, $E$, $X_1$, $X_2$, and $\xi$ be specified exactly as in simulation 3.1A of the main text. We relate the confounders to the potential outcomes in the following way: $\text{logit}(P(Y_i(1)=1))=\text{exp}(0.25X_{2i})+0.5X_{1i}X_{2i}$ and $\text{logit}(P(Y_i(0)=1))=0.2X_{2i}^3+0.25X_{1i}$. To obtain binary outcomes, we simulate $Y_i(1)\sim Bernoulli(P(Y_i(1)=1))$ and $Y_i(0)\sim Bernoulli(P(Y_i(0)=1))$. In 2B, $N$, $E$, $X_1$, $X_2$, and $\hat{\xi}$ are specified exactly as in simulation 3.1B of the main text, and we let $\text{logit}(P(Y_i(1)=1))=3(1+\exp(-8X_{2i}+1))^{-1}+0.25X_{1i}-2$ and $\text{logit}(P(Y_i(0)=1))=.85(X_{2i}-2)+(X_{2i}-2)^2$.

As in Section 3.1 of the main text, the resulting simulated datasets exhibit non-overlap in the right tail of the propensity score, and we can control the severity of the non-overlap through the $c$ parameter in the distribution of $X_2$. Again, we will ignore any small intervals of non-overlap (containing less than 10 observations) that may occur due to randomness outside the right tail of the propensity score. We consider $c=0$, $c=0.35$, and $c=0.7$, which yield data with properties similar to those of the simulations described in Section 3.1 of the main text. A sample dataset under each of these conditions is illustrated in Figure 4 in Section 3 of the Supplementary Materials.

We implement BART+SPL, untrimmed GR for binary outcomes (U-GR), and untrimmed BART probit (U-BART) on each of 1,000 simulated datasets under each condition. The results appear in Table~\ref{tab:results2}. All three methods perform similarly in simulation 2A; however, BART+SPL provides substantial gains in bias, coverage, and MSE in simulation 2B. We note that, even without non-overlap, BART probit can fail to provide improvements over parametric methods when sample sizes are small to moderate (i.e., we have found mediocre performance in some datasets with $N=500$), and thus we recommend that BART+SPL for binary outcomes only be applied to large datasets.

\begin{table}[h!]
\centering
\caption{Absolute (Abs) bias, 95\% credible interval coverage and mean square error (MSE) in estimation of the population average causal effects in simulations from Section 2.}
\begin{tabular}{crrrrr}
  \hline
Simulation Setting & Method & Abs Bias & Abs Bias (\%) & Coverage & MSE \\ 
  \hline
   \multirow{3}{*}{2Ai} & U-GR & 0.03 & 27.30 & 0.69 & 0.01 \\
  & U-BART & 0.03 & 25.73 & 0.94 & 0.01 \\
  & BART+SPL & 0.03 & 25.25 & 0.96 & 0.01 \\
  \hline
   \multirow{3}{*}{2Aii} & U-GR & 0.03 & 31.01 & 0.70 & 0.01 \\
  & U-BART & 0.03 & 28.17 & 0.97 & 0.01 \\
  & BART+SPL & 0.03 & 28.17 & 0.98 & 0.01 \\
  \hline
   \multirow{3}{*}{2Aiii} & U-GR & 0.04 & 38.88 & 0.67 & 0.01 \\
  & U-BART & 0.03 & 34.18 & 0.95 & 0.01 \\
  & BART+SPL & 0.03 & 35.25 & 0.98 & 0.01 \\
  \hline
 \multirow{3}{*}{2Bi} & U-GR & 0.07 & 81.83 & 0.38 & 0.10 \\
  & U-BART & 0.05 & 56.78 & 0.90 & 0.05 \\
  & BART+SPL & 0.04 & 47.50 & 0.95 & 0.04 \\
  \hline
   \multirow{3}{*}{2Bii} & U-GR & 0.08 & 102.09 & 0.35 & 0.10 \\
  & U-BART & 0.05 & 67.91 & 0.88 & 0.06 \\
  & BART+SPL & 0.04 & 54.03 & 0.98 & 0.04 \\
  \hline
   \multirow{3}{*}{2Biii} & U-GR &  0.10 & 132.34 & 0.28 & 0.10 \\
  & U-BART & 0.06 & 85.29 & 0.83 & 0.06 \\
  & BART+SPL & 0.05 & 62.46 & 0.99 & 0.04 \\
  \hline
\end{tabular}
\label{tab:results2}
\end{table}

\subsection{Supplementary Tables and Figures}
\label{as:suppfig}


\begin{figure}[H]
\begin{subfigure}[t]{1\textwidth}
\centering
\includegraphics[scale=.28]{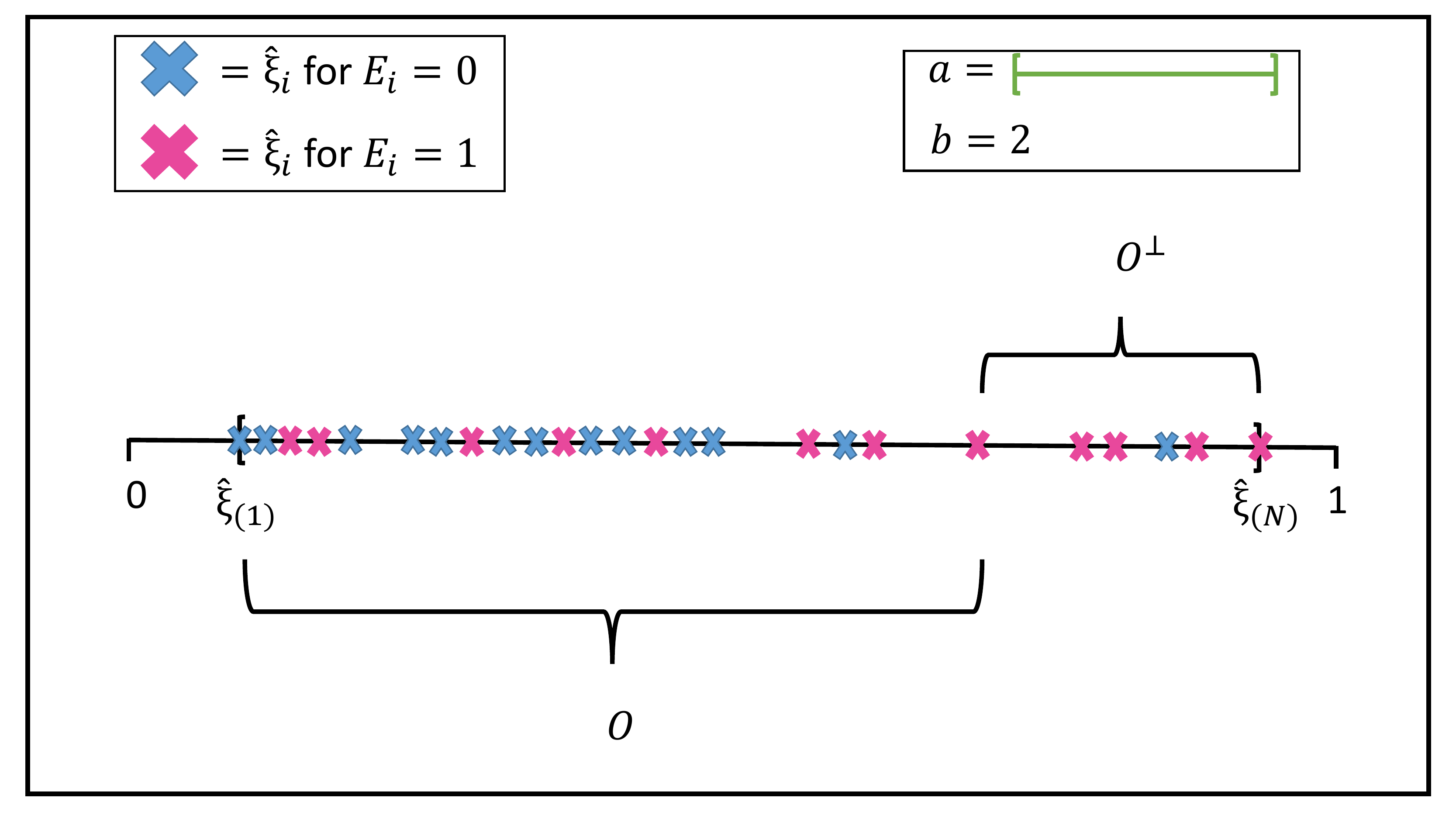}
\caption{}
\label{overlapA}
\end{subfigure}
\begin{subfigure}[t]{1\textwidth}
\centering
\includegraphics[scale=.28]{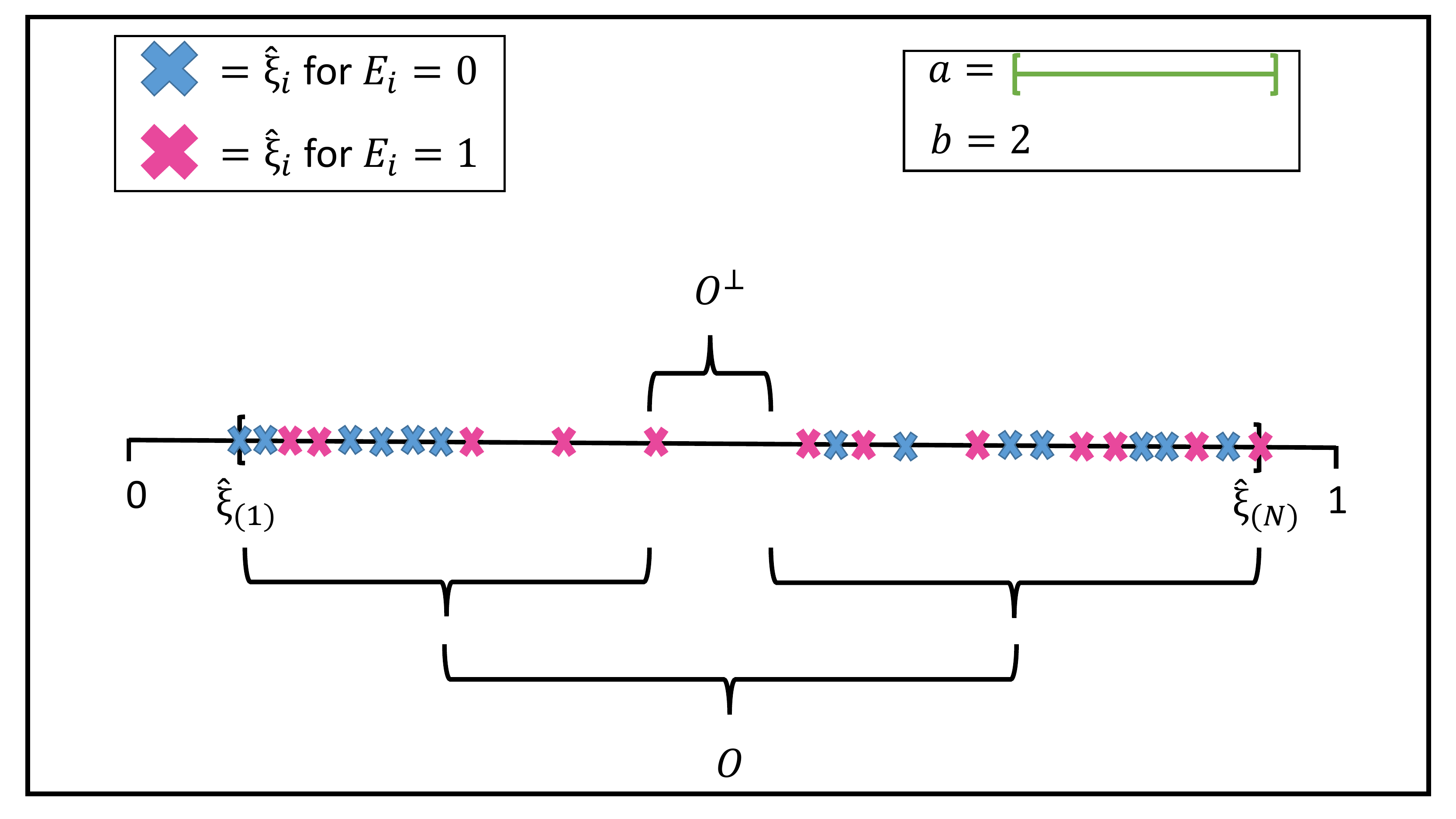}
\caption{}
\label{overlapB}
\end{subfigure}
\begin{subfigure}[t]{1\textwidth}
\centering
\includegraphics[scale=.28]{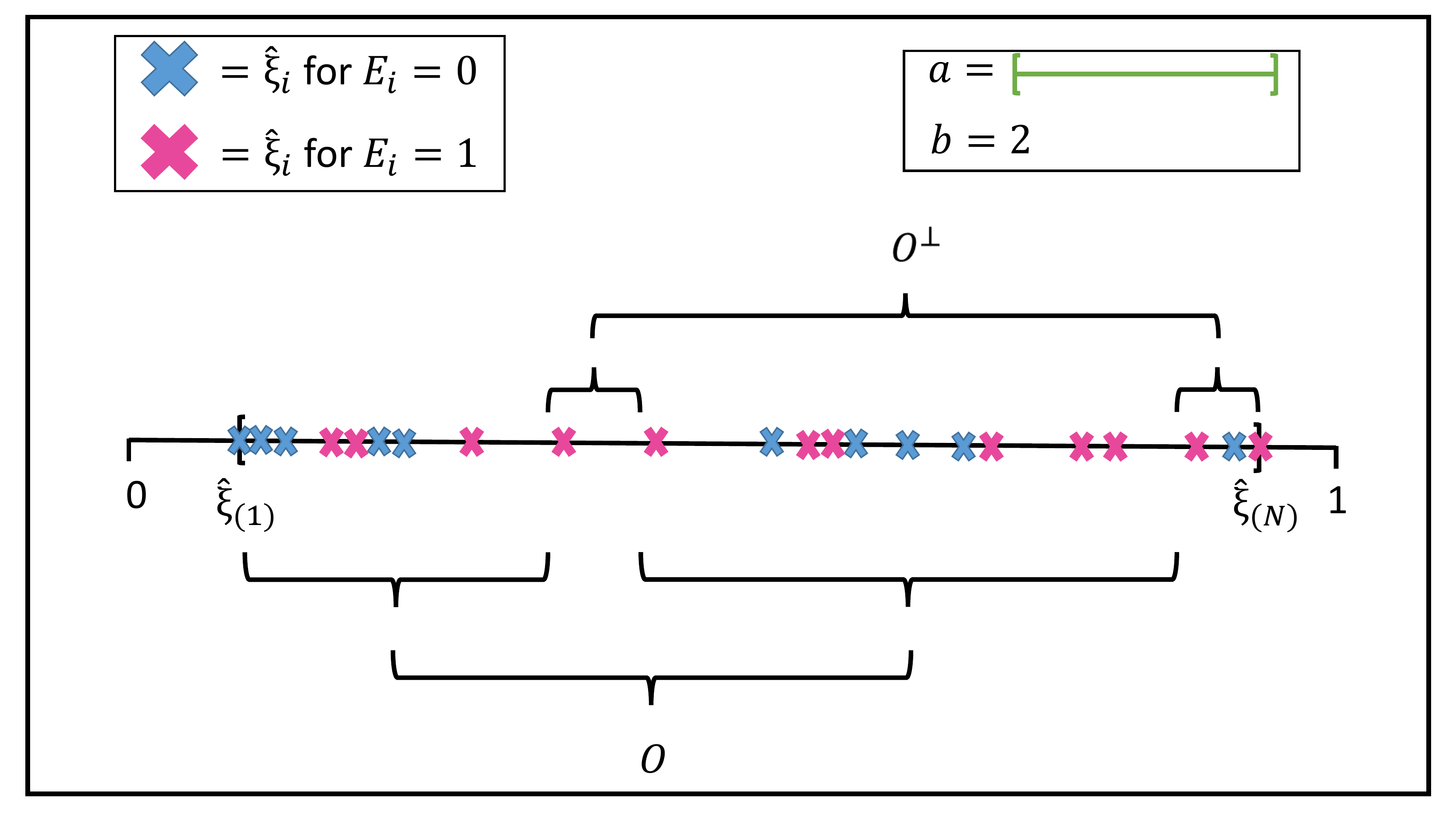}
\caption{}
\label{overlapC}
\end{subfigure}
\caption{Examples of several types of propensity score non-overlap regions that can be captured by our definition. The definition relies on user-specified parameters $a$ and $b$. $a$ is an interval length and $b$ is a portion of the sample size, representing the number of estimated propensity scores from each exposure group that must lie sufficiently close, i.e., within an interval of length $a$, to any given point in order for the point to be added to the RO. Figure (a) demonstrates non-overlap in the tail of the distribution, Figure (b) demonstrates non-overlap in the interior of the distribution, and Figure (c) demonstrates multiple intervals of non-overlap.}
\label{overlapall}
\end{figure}

\begin{landscape}
\begin{figure}[H]
\begin{subfigure}[t]{1\textwidth}
\centering
\includegraphics[scale=.54,page=4]{data_config_cont.pdf}~
\includegraphics[scale=.54,page=5]{data_config_cont.pdf}~
\includegraphics[scale=.54,page=6]{data_config_cont.pdf}
\caption{}
\label{fig:hista}
\end{subfigure}
\\
\begin{subfigure}[t]{1\textwidth}
\centering
\includegraphics[scale=.54,page=1]{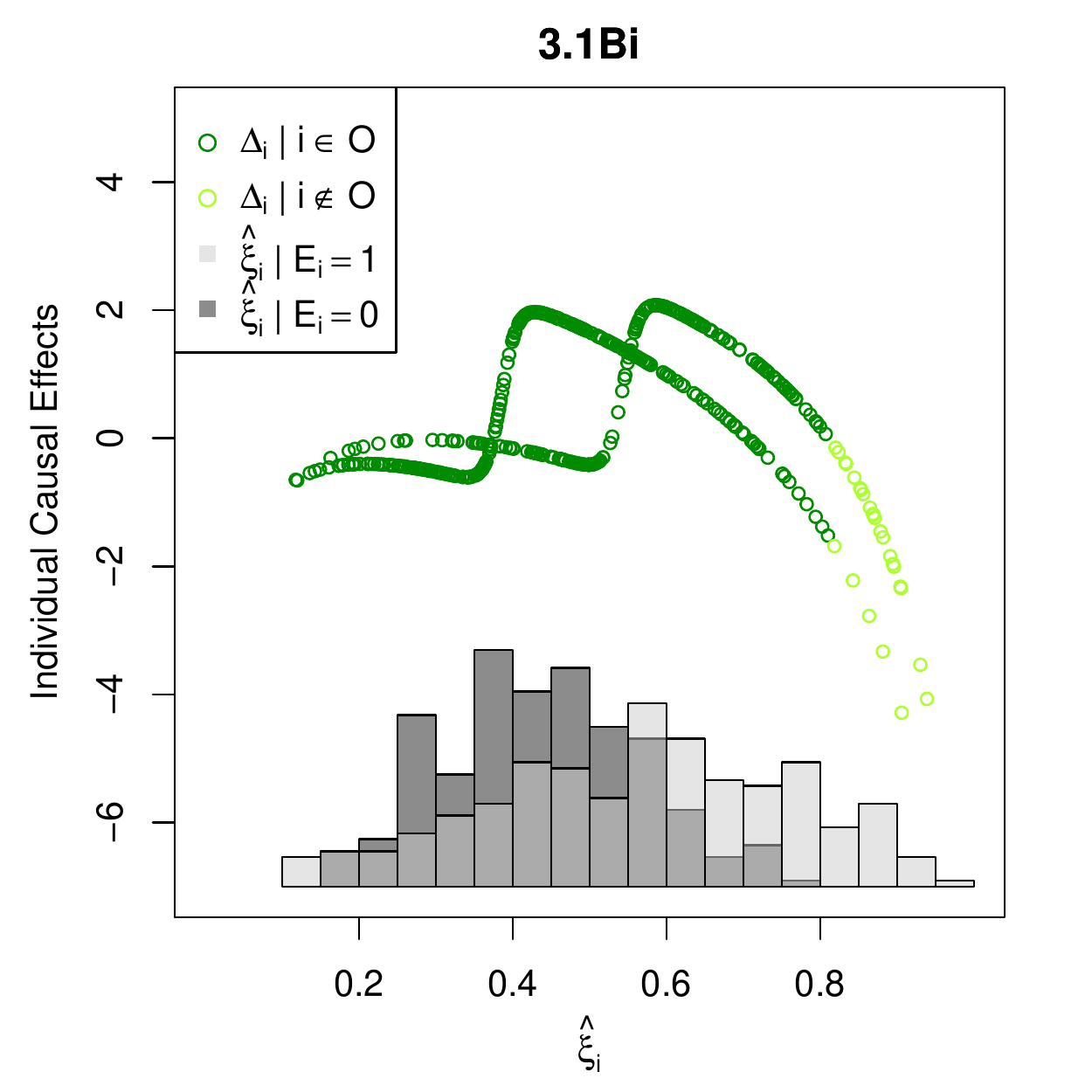}~
\includegraphics[scale=.54,page=2]{data_config_cont.pdf}~
\includegraphics[scale=.54,page=3]{data_config_cont.pdf}
\caption{}
\label{fig:histb}
\end{subfigure}
\caption{Examples of the simulated data for simulation 3.1A (a) and simulation 3.1B (b). The appearance of two separate trend lines in the figures is due to the use of one binary confounder and one continuous confounder for data generation.}
\label{fig:hist}
\end{figure}
\end{landscape}

\begin{landscape}
\begin{figure}[H]
\begin{subfigure}[t]{1\textwidth}
\centering
\includegraphics[scale=.7]{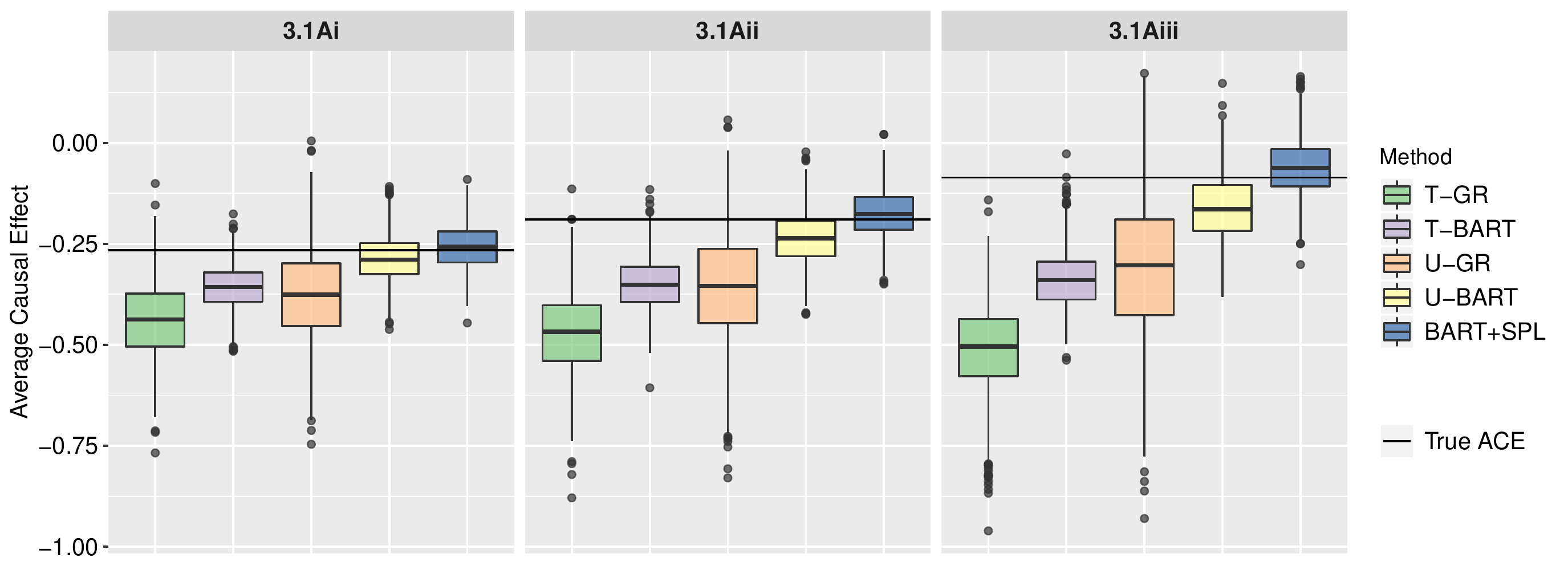}
\caption{}
\label{fig:resultsa}
\end{subfigure}
\\
\begin{subfigure}[t]{1\textwidth}
\centering
\includegraphics[scale=.7]{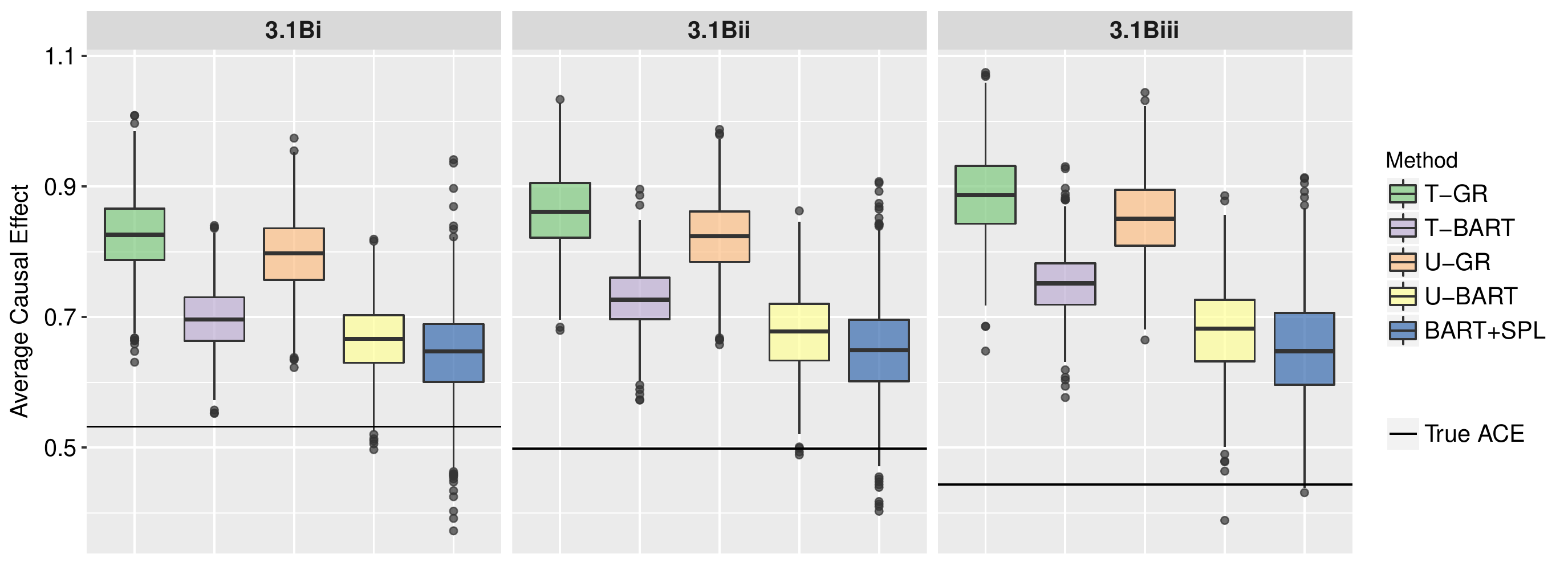}
\caption{}
\label{fig:resultsb}
\end{subfigure}
\caption{Estimates of Average Causal Effects from simulation 3.1A (a) and simulation 3.1B (b).}
\label{fig:results1}
\end{figure}
\end{landscape}

\begin{landscape}
\begin{figure}[H]
\begin{subfigure}[t]{1\textwidth}
\centering
\includegraphics[scale=.54,page=4]{data_config_bin.pdf}~
\includegraphics[scale=.54,page=5]{data_config_bin.pdf}~
\includegraphics[scale=.54,page=6]{data_config_bin.pdf}
\caption{}
\label{fig:histabin}
\end{subfigure}
\\
\begin{subfigure}[t]{1\textwidth}
\centering
\includegraphics[scale=.54,page=1]{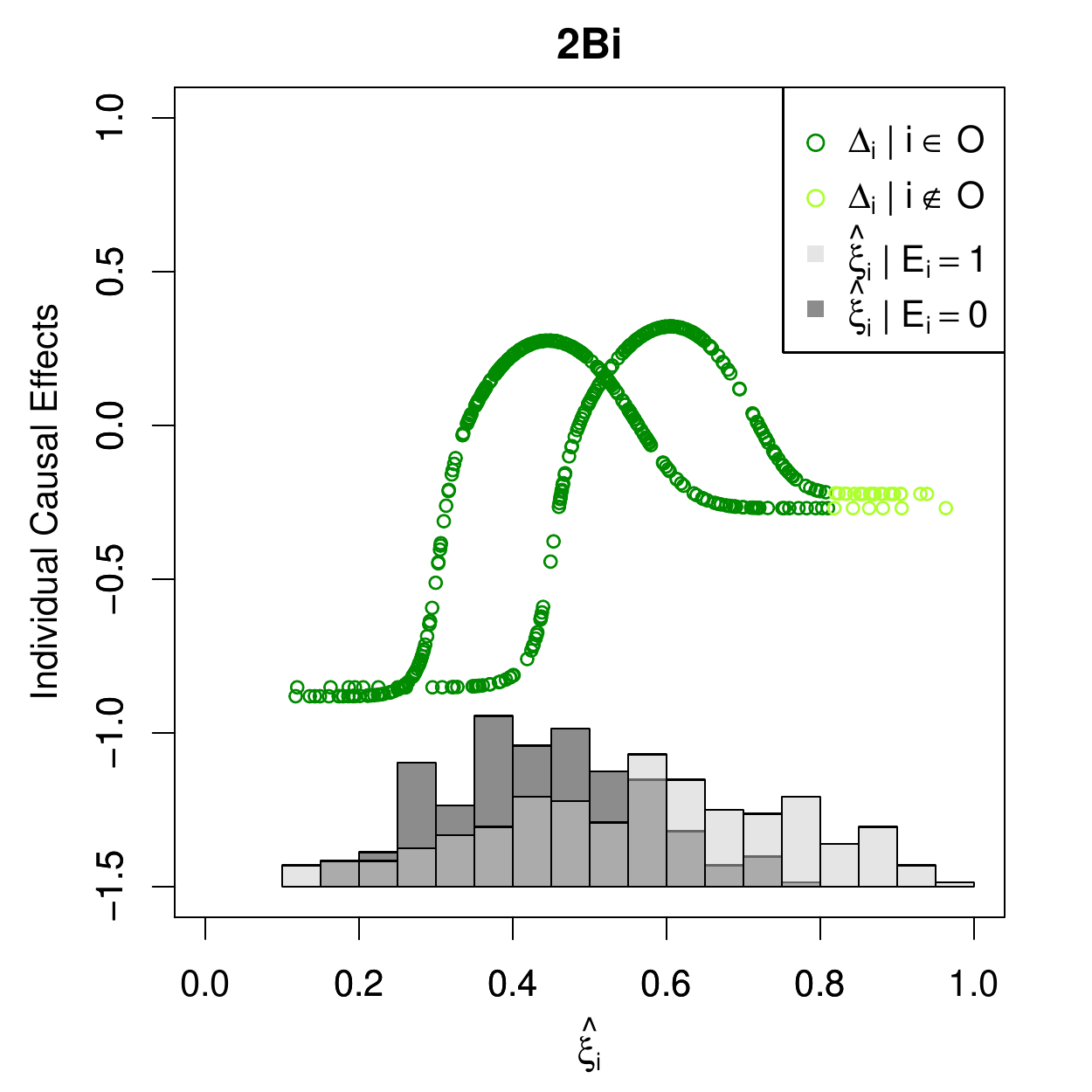}~
\includegraphics[scale=.54,page=2]{data_config_bin.pdf}~
\includegraphics[scale=.54,page=3]{data_config_bin.pdf}
\caption{}
\label{fig:histbbin}
\end{subfigure}
\caption{Examples of the simulated data for simulation 2A (a) and simulation 2B (b) of the Appendix. The appearance of two separate trend lines in the figures is due to the use of one binary confounder and one continuous confounder for data generation.}
\label{fig:histbin}
\end{figure}
\end{landscape}

\begin{figure}[H]
\centering
\includegraphics[scale=.68]{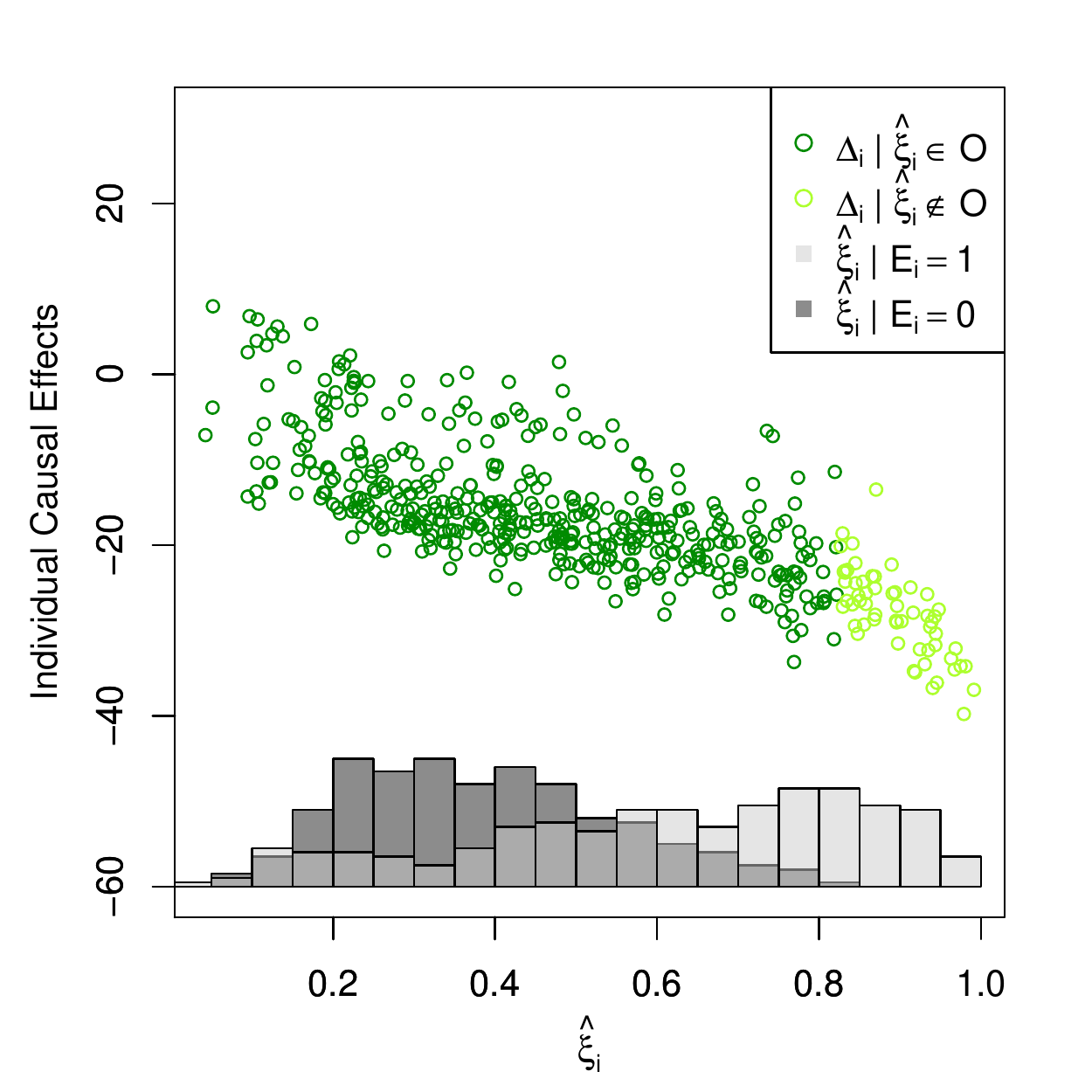}
\caption{Example dataset from simulation 3.2B where $p=35$, i.e., 10 true confounders and 25 randomly generated predictors.}
\label{fig:hd}
\end{figure}

\begin{figure}[H]
\begin{subfigure}[t]{.48\textwidth}
\centering
\includegraphics[scale=.68,page=1]{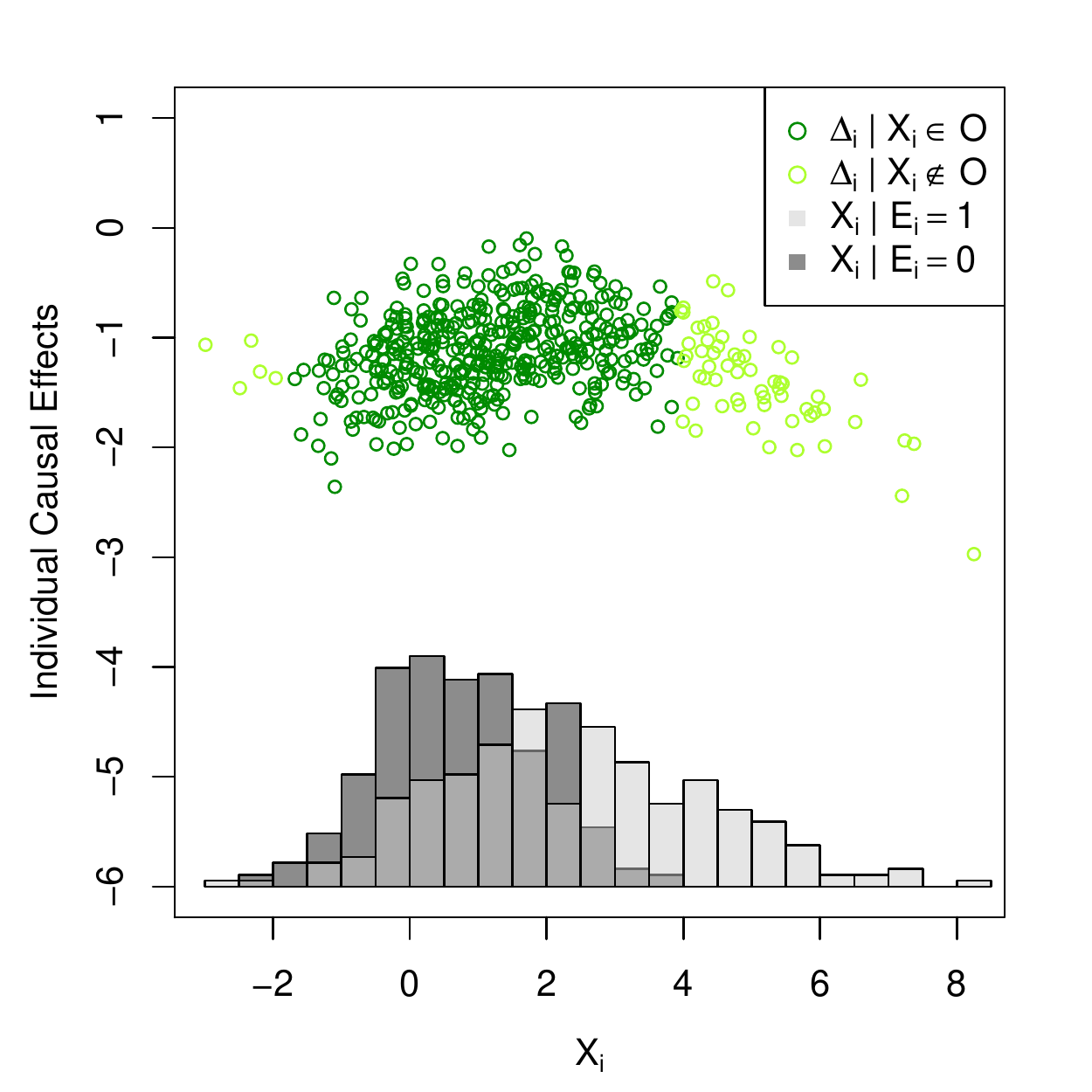}
\caption{}
\label{fig:easy}
\end{subfigure}
\begin{subfigure}[t]{.48\textwidth}
\centering
\includegraphics[scale=.68,page=2]{data_config_ab.pdf}
\caption{}
\label{fig:hard}
\end{subfigure}
\caption{Example datasets from simulation 3.3A (a) and simulation 3.3B (b).}
\label{fig:choosea}
\end{figure}

\begin{table}[H]
\centering
\caption{Summary of outcome and confounder variables overall and stratified by exposure status. All rates are per 100,000 population.}
\begin{tabular}{rrrrrrr}
& \multicolumn{2}{c}{Total (N=978)} & \multicolumn{2}{c}{Unexposed (N=687)} & \multicolumn{2}{c}{Exposed (N=291)}\\
\cline{2-3} \cline{4-5} \cline{6-7}
Variable & Mean & SD & Mean & SD & Mean & SD \\ 
  \hline
 2014 Thyroid Rate & 0.55 & 0.06 & 0.55 & 0.06 & 0.56 & 0.05 \\ 
 Change Thyroid Rate 1980-2014 & 1.71 & 8.03 & 1.16 & 7.67 & 2.99 & 8.71 \\ 
 2014 Leukemia Rate & 9.64 & 0.90 & 9.66 & 0.90 & 9.6 & 0.89 \\ 
 Change Leukemia Rate 1980-2014 & -4.29 & 8.54 & -4.26 & 8.24 & -4.36 & 9.21 \\ \hline
 Rate of Primary Care Physicians & 801.39 & 1330.09 & 773.65 & 1398.35 & 866.87 & 1152.66 \\ 
 Percent Uninsured & 17.74 & 5.53 & 17.36 & 5.48 & 18.64 & 5.55 \\ 
 Percent Diabetic & 83.02 & 8.14 & 83.33 & 8.07 & 82.29 & 8.29 \\ 
 Percent Current Smokers & 20.51 & 6.17 & 20.33 & 6.49 & 20.95 & 5.33 \\ 
 Percent Healthy Food Access & 10.18 & 8.19 & 10.33 & 8.67 & 9.82 & 6.95 \\ 
 Percent Obese & 28.43 & 5.19 & 28.45 & 5.25 & 28.40 & 5.05 \\ 
 Food Environment Index & 7.31 & 1.32 & 7.33 & 1.32 & 7.27 & 1.30 \\ 
 Population Density & 99.38 & 332.66 & 106.05 & 372.07 & 83.62 & 212.12 \\ 
 Percent Male & 50.10 & 1.89 & 50.10 & 1.83 & 50.11 & 2.05 \\ 
 Percent $<$55 & 69.61 & 6.17 & 68.88 & 6.46 & 71.36 & 5.01 \\ 
 Percent White & 84.88 & 16.39 & 86.16 & 15.77 & 81.86 & 17.42 \\ 
 Average Household Size & 2.50 & 0.25 & 2.47 & 0.25 & 2.56 & 0.26 \\ 
 Percent $\geq$ Bachelors Degree & 20.01 & 7.80 & 20.36 & 8.19 & 19.19 & 6.74 \\ 
 Percent Unemployed & 7.00 & 3.69 & 6.80 & 3.60 & 7.48 & 3.84 \\ 
  Median Household Income & 46296 & 10436 & 46100 & 10489 & 46759 & 10314 \\ 
  Gini Index of Inequality & 0.44 & 0.03 & 0.44 & 0.03 & 0.44 & 0.04 \\ 
  Percent Own Home & 71.85 & 7.44 & 72.00 & 7.63 & 71.48 & 6.97 \\ 
  Median Rent, Proportion Income & 27.43 & 4.43 & 27.28 & 4.41 & 27.76 & 4.45 \\ 
  Average Commute Time & 21.28 & 5.30 & 21.08 & 5.34 & 21.75 & 5.19 \\ 
   \hline
\end{tabular}
\label{tab:ngtab1}
\end{table}

\begin{table}[ht]
\centering
\caption{Average causal effects of natural gas compressor station presence on 2014 county-level thyroid cancer and leukemia mortality rates and the change in thyroid cancer and leukemia mortality rates from 1980 to 2014. Sensitivity analysis with $a=.05*range(\hat{\xi})$ and $b=10$.}
\begin{tabular}{rrrr}
  \hline
Outcome & Method & Effect & 95\% CI \\ 
  \hline
\multirow{2}{*}{2014 Thyroid Rates} & BART+SPL & -0.001 & -0.030, 0.028 \\ 
									& BART & 0.002 & -0.008, 0.011 \\ \hline
\multirow{2}{*}{Change in Thyroid Rates 1980-2014} & BART+SPL & 1.020 & -0.531, 2.616 \\ 
												   & BART & 1.039 & 0.050, 2.019 \\ \hline
\multirow{2}{*}{2014 Leukemia Rates} & BART+SPL & 0.003 & -0.029, 0.034 \\ 
									 & BART & 0.005 & -0.004, 0.015 \\ \hline
\multirow{2}{*}{Change in Leukemia Rates 1980-2014} & BART+SPL & 0.963 & -0.655, 2.557 \\
													& BART & 0.886 & -0.140, 1.909 \\ 
   \hline
\end{tabular}
\label{sens1}
\end{table}

\begin{table}[ht]
\centering
\caption{Average causal effects of natural gas compressor station presence on 2014 county-level thyroid cancer and leukemia mortality rates and the change in thyroid cancer and leukemia mortality rates from 1980 to 2014. Sensitivity analysis with $a=.15*range(\hat{\xi})$ and $b=3$.}
\begin{tabular}{rrrr}
  \hline
Outcome & Method & Effect & 95\% CI \\ 
  \hline
\multirow{2}{*}{2014 Thyroid Rates} & BART+SPL & 0.003 & -0.008, 0.014  \\ 
									& BART & 0.002 & -0.007, 0.012  \\ \hline
\multirow{2}{*}{Change in Thyroid Rates 1980-2014} & BART+SPL & 0.980 & -0.141, 2.127  \\ 
												   & BART & 0.939 & -0.047, 1.884  \\ \hline
\multirow{2}{*}{2014 Leukemia Rates} & BART+SPL & 0.003 & -0.008, 0.015 \\ 
									 & BART & 0.002 & -0.007, 0.012 \\ \hline
\multirow{2}{*}{Change in Leukemia Rates 1980-2014} & BART+SPL & 0.851 & -0.316, 2.001 \\
													& BART & 0.864 & -0.110, 1.836 \\ 
   \hline
\end{tabular}
\label{sens2}
\end{table}

\end{document}